\documentclass[journal,10pt]{IEEEtran}
\makeatletter
\def\endthebibliography{%
    \def\@noitemerr{\@latex@warning{Empty `thebibliography' environment}}%
    \endlist
}
\makeatother

\usepackage{cite}

\usepackage[pdftex]{graphicx}
\usepackage[caption=false,font=footnotesize]{subfig}

\usepackage{amsmath}
\usepackage{mathtools, cuted}
\usepackage{amssymb}
\usepackage{bm}
\usepackage{mathrsfs}
\usepackage{breqn}
\usepackage{booktabs}

\usepackage{pifont}
\usepackage{xcolor}
\usepackage{url}
\usepackage{lettrine} 
\usepackage{lipsum}
\usepackage{siunitx}
\usepackage{soul}
\usepackage{array}
\usepackage[inline]{enumitem}
\usepackage{epsfig}
\usepackage{filecontents}

\usepackage{algpseudocode}
\usepackage{algorithm, tabularx}
\usepackage{multirow}
\newcolumntype{L}[1]{>{\raggedright\let\newline\\\arraybackslash\hspace{0pt}}m{#1}}
\newcolumntype{C}[1]{>{\centering\let\newline\\\arraybackslash\hspace{0pt}}m{#1}}
\newcolumntype{R}[1]{>{\raggedleft\let\newline\\\arraybackslash\hspace{0pt}}m{#1}}
\newlength{\maxwidth}

\makeatletter
\newcommand{\multiline}[1]{%
	\begin{tabularx}{\dimexpr\linewidth-\ALG@thistlm}[t]{@{}X@{}}
		#1
	\end{tabularx}
}
\makeatother
\algdef{SE}[SUBALG]{Indent}{EndIndent}{}{{\algorithmicend\ }}
\algtext*{Indent}
\algtext*{EndIndent}

\usepackage{amsthm}

\theoremstyle{remark}

\begin{document}

\title{Unlocking Downlink NOMA with FARIS:\\Joint Clustering and Surface Configuration Design}

\author{Hong-Bae Jeon,~\IEEEmembership{Member,~IEEE}, and Tuo Wu
\thanks{\textit{(Corresponding Author: Hong-Bae Jeon)}}%
\thanks{H.-B. Jeon is with the School of Electronic Engineering, Soongsil University, Seoul, Korea (e-mail: hongbae08@ssu.ac.kr).}%
\thanks{T. Wu is with the School of Electronic and Information Engineering, South China University of Technology, Guangzhou, China (e-mail:wutuo@scut.edu.cn).}
}

\maketitle

\begin{abstract}
This paper investigates a fluid active reconfigurable intelligent surface (FARIS)-aided downlink non-orthogonal multiple access (NOMA) system. We formulate a network sum-rate maximization problem that jointly optimizes user clustering, NOMA power allocation, FARIS amplification gains, discrete phase shifts, and fluid element selection under quality-of-service, reflected-power, and hardware constraints. To address the resulting nonconvex mixed-integer problem, we develop a two-stage framework comprising distance-based interleaved clustering for constructing successive-interference-cancellation (SIC)-friendly user groups and per-cluster alternating optimization. The resulting subproblems are handled using geometric programming (GP), fractional programming (FP), majorization-minimization (MM) with mixed-integer phase optimization, and the cross-entropy method (CEM). Numerical results demonstrate rapid convergence, near-optimal performance relative to brute-force search (BFS)-based optimum, and consistently outperforms the benchmarks. These results verify the effectiveness of jointly integrating FARIS and NOMA for high-rate downlink transmission.
\end{abstract}


\IEEEpeerreviewmaketitle  

\section{Introduction}
\label{sec:intro}
\lettrine{T}{he} vision of reconfigurable intelligent surface (RIS) lies in transforming the wireless propagation environment from an uncontrollable medium into a programmable entity~\cite{RIST,risspm}. By properly coordinating the phase responses of individual elements, RIS can flexibly manipulate the propagation of incident signals, thereby enhancing coverage, link reliability, spectral efficiency, and localization performance across diverse scenarios~\cite{nfRIS, HBRIS, HBRIS22}. However, despite this promising paradigm, RIS is fundamentally constrained by the multiplicative fading effect, where the cascaded channel suffers from severe multiplicative path-loss~\cite{vsrelay}. This intrinsic limitation critically undermines the achievable signal strength and coverage, particularly in blockage-dominated scenarios, and has emerged as a key bottleneck in realizing the full potential of RIS-aided systems~\cite{RISnearmag}. To overcome this fundamental challenge, active-RIS (ARIS) has recently been proposed as a compelling architectural evolution~\cite{aristut, aris1}.

Unlike conventional passive-RIS (PRIS), ARIS equips each reflecting element with a reflection-type amplifier, enabling the surface to actively amplify the incident signal before re-radiation~\cite{aris5}. This capability allows ARIS to effectively compensate for the severe cascaded path loss and mitigate the multiplicative fading effect. Prior works have demonstrated that ARIS can significantly enhance the system performance under various settings, including joint beamforming optimization with sub-connected architectures~\cite{aris2, arissub22}, signal-to-noise-ratio (SNR)-oriented analyses under identical power budget with PRIS~\cite{aris4, aris5}, integration with integrated-sensing-and-communications (ISAC)~\cite{crbaris, arissub22} and high-altitude-platform (HAP)~\cite{HBRIS22, arishap, arisjsachap}, average sum-rate maximization~\cite{aris8} and power-minimizing~\cite{aris7} under partial channel-state-information (CSI) conditions, and finding its linear-time optimal configuration~\cite{Doh}, highlighting its potential to offer greater flexibility than PRIS in shaping the end-to-end performance.

However, despite these advantages, ARIS still relies on a fixed and static surface geometry, inherited from conventional RIS architectures. In practice, the reflecting elements are confined to predetermined positions with discretized phase control, which fundamentally limits the achievable spatial adaptability of the surface. This limitation becomes particularly critical when aiming to exploit higher spatial degree-of-freedom (DoF), as substantial DoF gains typically require increasing the number of elements~\cite{aris1}, leading to higher training overhead and implementation complexity. As a result, even with active amplification, the performance of ARIS remains constrained by its inability to adaptively exploit favorable spatial locations.

To overcome this structural limitation, the concept of fluid antenna system (FAS) has recently been introduced~\cite{FAS, wu11}, enabling dynamic repositioning of antenna elements within a given region, thereby introducing additional physical-layer DoFs that can be exploited for performance enhancement~\cite{fasopdg, vcfas22}. Since its introduction, a growing body of research has investigated various aspects of FAS, including its fundamental performance limits~\cite{fasopdg, FASqout}, channel estimation techniques tailored to fluid architectures~\cite{fasheath, fasover}, and its integration into a wide range of applications such as integration with ISAC~\cite{fasisac, fasisac22} and RIS~\cite{FASARISHB, fasrisper, FASRISsec}, and direction-of-arrival (DoA) estimation~\cite{fasdoa, fasdoa22}. Collectively, these studies establish FAS as a promising paradigm that goes beyond conventional fixed-structure antenna systems, by unlocking enhanced spatial diversity and offering a flexible platform that can be seamlessly incorporated into various 6G technologies.

By leveraging this idea, fluid RIS (FRIS) extends spatial reconfigurability to intelligent surfaces~\cite{FRISlook, FRISmag}, allowing a small number of movable elements to traverse a larger aperture and select advantageous propagation points. Through such fluidity, FRIS can effectively extract spatial-domain DoFs without increasing the number of active elements, providing a scalable and hardware-efficient approach to performance enhancement~\cite{FRISpa, FRISonoff}. Early studies on FRIS have established its fundamental performance advantages over conventional RIS architectures. In particular,~\cite{FRISlook} and~\cite{FRISpa} introduced the concept of position reconfigurability and developed corresponding analytical and optimization frameworks for FRIS-assisted systems. Specifically,~\cite{FRISlook} demonstrated substantial performance gains over conventional RIS in both single- and multi-user scenarios through joint port-selection and phase optimization, whereas~\cite{FRISpa} provided rigorous analytical characterizations of outage probability and capacity, including tight upper-bounds under statistical channel models. From a practical perspective,~\cite{FRISonoff} investigated implementable FRIS architectures with discrete phase shifts and port selection, showing that significant performance improvements can be achieved with reduced hardware complexity. Building upon these foundations, subsequent works have extended FRIS to a variety of application scenarios, highlighting its effectiveness in alleviating fundamental propagation limitations. In particular, the potential of FRIS for secure communications was explored in~\cite{FRISsec} and~\cite{FRISsec22}, where dynamic port selection was leveraged to enhance secrecy performance by exploiting spatial correlation, thereby strengthening legitimate links while suppressing eavesdropping channels. In~\cite{FRISim}, the authors proposed a FRIS-enabled index modulation that leverages fluid element repositioning and phase control to encode information in spatial indices, demonstrating the potential of FRIS as an additional information-bearing dimension. More recently in~\cite{FRISbp}, the authors extended the FRIS framework by introducing element-level radiation pattern reconfigurability and developing a joint beamforming and pattern co-design scheme to further enhance spatial signal control.

These results collectively demonstrate that FRIS provides an additional spatial DoF beyond conventional phase-only control, significantly improving link reliability, achievable rate, and security performance. Nevertheless, FRIS fundamentally relies on passive reflection and therefore cannot overcome the inherent multiplicative fading effect. This reveals a critical gap between \textbf{signal amplification} and \textit{\textbf{spatial adaptability}}, which are separately addressed by ARIS and FRIS, respectively. In particular, ARIS determines \textbf{how strongly} the signal is reinforced but lacks spatial flexibility, whereas FRIS determines \textit{\textbf{where}} the signal interaction occurs but lacks the ability to sufficiently boost the signal power.

Motivated by this observation, we argue that a deeper integration of these two paradigms can unlock a fundamentally new design space. Specifically, if each reflecting element is endowed with both spatial mobility and active amplification capability, the surface can jointly optimize \textit{\textbf{where}} the signal is reflected and \textbf{how strongly} it is reinforced. This leads to a new architecture, referred to as fluid-active-RIS (FARIS), proposed by \textit{\textbf{Jeon}}~\cite{FARIS}, in which each element operates as a fluidic active unit capable of dynamic repositioning, controllable amplification, and phase adjustment. Such an intrinsically integrated design creates a coupled spatial-amplitude optimization space that is not available in conventional A/FRIS systems. By jointly adapting the spatial distribution and the strength of the reflected signals, FARIS enables more efficient exploitation of spatial and power resources, thereby effectively addressing both multiplicative fading and the limitations in spatial DoF of conventional architectures.

A particularly compelling and natural application of FARIS emerges in non-orthogonal multiple access (NOMA) systems~\cite{nomawm}. By enabling multiple users to share the same time-frequency resources through power-domain multiplexing, NOMA can significantly improve spectral efficiency compared to conventional orthogonal multiple access (OMA)~\cite{fasnoma, firesnoma}. However, its performance critically depends on the existence of sufficient channel disparity among users to ensure effective successive interference cancellation (SIC)~\cite{FASnoma11, FASnoma22}. In practice, such disparity is often limited by the static nature of conventional propagation environments, thereby constraining the achievable performance gains of NOMA. In this regard, FARIS provides a unique opportunity to fundamentally enhance NOMA systems. Specifically, the fluid reconfigurability of FARIS allows the surface to \textbf{\textit{dynamically select spatial locations}} that induce favorable channel disparities among users, while its \textbf{active reflection capability} can further reinforce the effective cascaded channel gains. This dual capability enables FARIS to jointly control both the {relative channel ordering} and the {absolute signal strength}, which are two key factors governing the performance of NOMA~\cite{dingnoma, jingnoma}. As a result, FARIS can naturally facilitate more efficient SIC and improve overall system throughput by creating a more favorable propagation environment tailored for NOMA transmission.

Despite this strong synergy, the integration of FARIS and NOMA remains largely unexplored, where even the closely related line of research on FRIS-assisted NOMA systems is still in its infancy. Although a few recent studies have begun to investigate FRIS-enabled NOMA frameworks in both uplink~\cite{FRISnomaw} and downlink settings~\cite{FRISnoma}, these studies are fundamentally limited by the passive nature of FRIS, which cannot compensate for the severe multiplicative fading effect inherent in RIS-assisted channels. As a result, the achievable performance gains rely solely on spatial reconfigurability without addressing the underlying signal attenuation bottleneck, leaving significant performance headroom. The limitation holds vice versa for ARIS-assisted NOMA, which mitigates attenuation but lacks spatial adaptability. Thereby, to the best of our knowledge, the joint exploitation of spatial mobility and active reflection for NOMA has not yet been fully understood, and the corresponding system design and optimization are not investigated. In particular, the tight coupling between NOMA operation and FARIS design introduces significant challenges that remain largely unaddressed in the existing literature.

Motivated by these observations, this paper investigates a FARIS-aided downlink NOMA system, where the spatial-amplitude coupled design of FARIS is fully leveraged to enhance spectral efficiency. In particular, by integrating FARIS with power-domain multiplexing, the proposed design effectively enhances the cascaded channel gain while inducing favorable channel disparity among users, thereby enabling efficient SIC and improved sum-rate performance. By jointly optimizing user clustering, power allocation, and FARIS configuration, the proposed framework effectively exploits both the spatial flexibility and amplification capability of FARIS, thereby achieving substantial performance gains over conventional RIS-assisted and existing baseline schemes. The main contributions of this paper are summarized as follows:
\begin{itemize}
\item 
We first develop a system model for a FARIS-aided downlink NOMA network, where the fluid spatial mobility and active reflection capability of FARIS are jointly incorporated into the cascaded base-station (BS)-FARIS-user channels. In the proposed model, the FARIS simultaneously determines the spatial positions of the selected reflecting elements, their active reflection gains, and their discrete phase shifts, while the BS serves multiple users via power-domain superposition coding. This leads to a new coupled spatial-amplitude signal model in which the effective channel gains, user ordering for SIC, amplifier-noise propagation, and reflected-power consumption are all jointly governed by the FARIS configuration. As such, the proposed model captures a fundamentally richer interaction between surface reconfigurability and NOMA transmission than conventional A/FRIS-assisted systems.

\item 
Based on the proposed system, we formulate a network-wide sum-rate maximization problem for FARIS-aided downlink NOMA by jointly optimizing user clustering, NOMA power allocation, active reflection gains, discrete phase shifts, and fluid element selection. The formulation explicitly incorporates the reflected-power constraint of FARIS, amplifier noise, hardware power consumption, discrete phase quantization, binary on-off element selection, and minimum rate requirements for all users. The resulting problem is a highly nonconvex mixed-integer nonlinear program (MINLP), in which the NOMA and FARIS configuration are tightly coupled through the system design variables.

\item 
To solve the problem, we develop a two-stage framework consisting of distance-based interleaved NOMA clustering and per-cluster alternating optimization (AO). In the first stage, we propose a lightweight clustering strategy that exploits large-scale channel disparity to construct SIC-friendly NOMA groups with negligible overhead. In the second stage, we develop a structured AO procedure that decomposes the original intractable problem into four tractable blocks: i) a geometric-programming (GP)-based update for NOMA power allocation, ii) a fractional-programming (FP)-based amplification-gain design using the Lagrangian dual and quadratic transforms, iii) a majorization-minimization (MM)-based phase optimization followed by mixed-integer linear program (MILP), and iv) a cross-entropy method (CEM)-based element-selection update. The resulting framework preserves tractability while generating a nondecreasing objective sequence across iterations.

\item 
We provide extensive simulations to validate the proposed FARIS-NOMA design under various system parameters. The results demonstrate rapid and stable convergence of the proposed AO framework, and further show that the obtained solution achieves near-optimal performance compared with brute-force search (BFS) method. Moreover, the proposed FARIS-NOMA consistently outperforms FARIS-OMA, FRIS-NOMA, and ARIS-NOMA benchmarks, verifying that the joint exploitation of fluid reconfiguration, active reflection, and power-domain multiplexing provides substantial gains in network and average cluster sum-rate.
\end{itemize}

\begin{figure}[t]
	\begin{center}
		\includegraphics[width=0.8\columnwidth,keepaspectratio]%
		{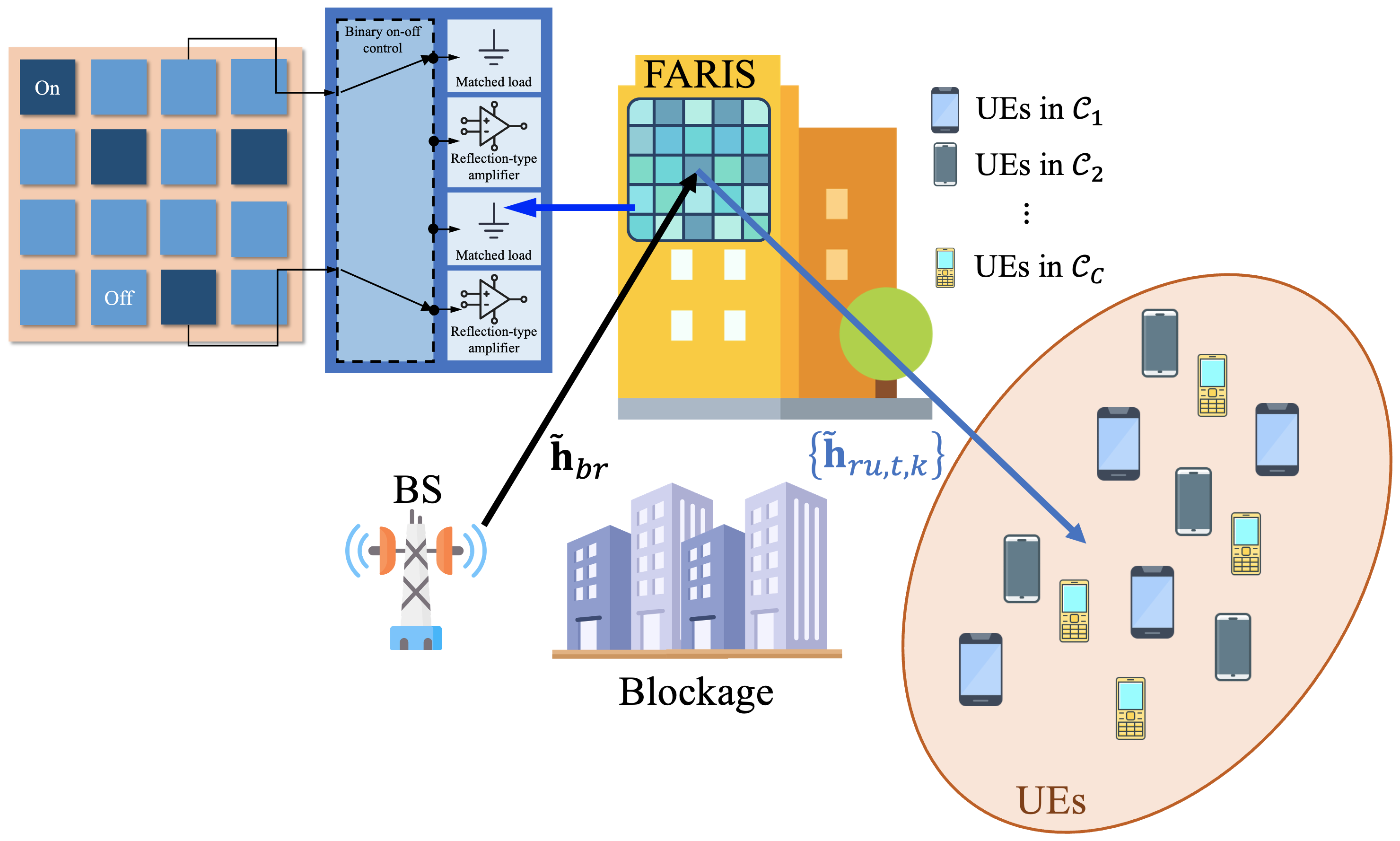}
		\caption{System model of FARIS-aided downlink NOMA systems.}
		\label{fig_sys}
	\end{center}
\end{figure}

\section{System Model}
\subsection{NOMA-Based Transmission with Multiple User Clusters}
We consider a FARIS-assisted downlink NOMA system composed of a BS, a FARIS, and $U$ single fixed-position
antenna (FPA) user equipments (UEs), indexed by $\mathcal U=\{1,\cdots,U\}$, as illustrated in Fig.~\ref{fig_sys}. The direct BS-UE links are assumed to be blocked~\cite{FRISnoma, FRISonoff}, such that all downlink transmissions are established through the FARIS. The UEs are partitioned into $C$ clusters, denoted by $\{\mathcal C_t\}_{t=1}^{C}$ with $K_t=|\mathcal C_t|$, where each cluster is served via power-domain NOMA within an orthogonal time slot, thereby avoiding inter-cluster interference, while intra-cluster interference is mitigated by means of SIC~\cite{FRISnoma, iotnoma}. Since different clusters operate over orthogonal slots, the analysis can be restricted to an arbitrary $C_t$ without loss of generality. The cluster $C_t$ is defined as $C_t \triangleq \{u \in \mathcal U : \xi_t^k = 1\}$, where $\xi_t^k=1$ indicates that $\mathrm{UE}_u$ belongs to $C_t$, and $\xi_t^k=0$ otherwise. 

\subsection{Signal Model of FARIS-Aided Downlink NOMA}
The FARIS consists of $N=N_x\times N_x$ candidates of fluid reflecting elements arranged over a planar aperture of size $W_x \lambda\times W_x \lambda$, where $\lambda$ is the carrier wavelength, and $N_o(\le N)$ elements are selected. The inter-element spacing is therefore given by $d_x=\frac{W_x\lambda}{N_x}$. Due to the dense element placement, the spatial correlation across the surface is captured by the matrix $\mathbf \Psi\in\mathbb C^{N\times N}$, whose $(p,q)$th entry follows the Jakes model: $[\mathbf \Psi]_{p,q}=J_0 \left(\frac{2\pi d_{p,q}}{\lambda}\right)$. Following~\cite{FRISonoff, FRISmag}\footnote{Although originally developed for FRIS, this model is directly applicable to FARIS~\cite{FARIS} due to their shared hardware architecture, with the distinction being the additional amplification capability in FARIS.} and as illustrated in Fig.~\ref{fig_sys}, each FARIS element is modeled as a port within a fluid antenna structure~\cite{FAS} and operates in a binary mode. Specifically, when in the \textbf{on} state, the element actively interacts with the incident electromagnetic wave, enabling controllable amplitude and phase adjustments. In contrast, when in the \textbf{off} state, the element is terminated with a matched load, thereby effectively isolating it from the impinging signal and suppressing any reflection. Based on this hardware structure, we later develop a corresponding power consumption model along with the associated reflected-power constraints.

Let $\mathbf h_{br},\mathbf h_{ru,t,k}\sim \mathcal{CN}(\mathbf 0,\mathbf I_N)$ denote the small-scale BS-to-FARIS and FARIS-to-$\mathrm{UE}_{t,k}$ (UE $k$ in cluster $t$) fadings, respectively~\cite{FRISnoma}. The correlated channels are then defined as $\tilde{\mathbf h}_{br}=\mathbf \Psi^{1/2}\mathbf h_{br}$ and $\tilde{\mathbf h}_{ru,t,k}=\mathbf \Psi^{1/2}\mathbf h_{ru,t,k}$, and the corresponding large-scale fadings are $ B_{br}=\tau d_{br}^{-\zeta},~ B_{ru,t,k}=\tau d_{ru,t,k}^{-\zeta}$ and $B_{t,k}=B_{br}B_{ru,t,k}$, where $\zeta$ is the PL exponent (PLE), $\tau$ is the reference path loss (PL) at a distance of 1~m, and $d_{br}$ and $d_{ru,t,k}$ are the distances for the BS-FARIS and FARIS-$\mathrm{UE}_{t,k}$, respectively. Let $\mathbf D_{\chi}=\mathrm{diag}(\chi^{(1)},\cdots,\chi^{(N)})$ denote the element-selection matrix with $\chi^{(n)}\in\{0,1\}$ and $\sum_{n=1}^{N}\chi^{(n)}=N_o$. The operation at FARIS is therefore: 
\begin{equation}
\label{gt}
\mathbf \Gamma \triangleq \mathbf D_{\chi}\mathbf A\mathbf \Phi,
\end{equation}
where $\mathbf A=\mathrm{diag}(\alpha_1,\cdots,\alpha_N),~\mathbf \Phi=\mathrm{diag}(e^{j\theta_1},\cdots,e^{j\theta_N})$, with $\alpha_n\in [0, \alpha_{\max}]$ denoting the active reflection gain and $\theta_n\in\mathcal Q_b\triangleq\{\frac{2\pi i}{2^b} : i=0, \cdots, 2^b -1\}$ denoting the discrete phase shift of the $n$th element with $b$ bit discretization.

In slot $t$, the BS transmits the superposition-coded: $x_t=\sum_{k\in\mathcal C_t}\sqrt{\rho_{t,k}P_t} x_{t,k}$, where $\mathbb E[|x_{t,k}|^2]=1$, $P_t$ is the BS power allocated to $\mathcal C_t$, and the power allocation coefficient $\{\rho_{t,k}\}$ along $\mathcal C_t$ satisfies $\sum_{k\in\mathcal C_t}\rho_{t,k}=1$, and $0< \rho_{t,k}\le 1$. The signal incident on the FARIS is therefore
\begin{equation}
\label{sif}
\mathbf r_t=\sqrt{B_{br}} \tilde{\mathbf h}_{br}x_t+\mathbf n_{F,t},
\end{equation}
where $\mathbf n_{F,t}\sim\mathcal{CN}(\mathbf 0,\sigma_F^2\mathbf I_N)$ models the amplifier noise introduced by FARIS. Accordingly, the reflected signal is
\begin{equation}
\label{rsd}
\mathbf s_{F,t}=\mathbf \Gamma \mathbf r_t=\mathbf D_{\chi}\mathbf A\mathbf \Phi\left(\sqrt{B_{br}} \tilde{\mathbf h}_{br}x_t+\mathbf n_{F,t}\right).
\end{equation}
Due to the active reflection mechanism, the FARIS is subject to a total reflected-power constraint. Specifically, the transmit power radiated by the FARIS in slot $t$ is given by $P_{F,t}=\mathbb E\left[\|\mathbf s_{F,t}\|^2\right]$. Substituting~\eqref{rsd} and using $\mathbb E[|x_t|^2]=P_t$:
\begin{equation}
\label{pft}
P_{F,t}
=
B_{br}P_t\left\|\mathbf D_{\chi}\mathbf A\mathbf \Phi\tilde{\mathbf h}_{br}\right\|_2^2
+
\sigma_F^2\left\|\mathbf D_{\chi}\mathbf A\mathbf \Phi\right\|_F^2,
\end{equation}
which is independent of \(\mathbf\Phi\) and accounts for both the amplified incident signal component and the amplified thermal noise emitted by the FARIS. Beyond $P_{F,t}$, the FARIS architecture also introduces hardware power consumption associated with the fluid-active ports. Specifically, following the hardware structure and circuit model illustrated in Fig.~\ref{fig_sys}~\cite{FARIS}, the circuit power consumption of FARIS consists of two components: i) $P_c$, representing the logical control and switching power required for each candidate element, and ii) $P_{\mathrm{DC}}$, representing the direct current (DC) bias power needed to support active reflection~\cite{aris5}. Since the logical control circuitry is implemented for all $N$ candidate ports, whereas the active reflection branch is activated only for the selected $N_o$ FARIS ports, the total power consumption of FARIS can be expressed as
\begin{equation}
\label{eq:Pfaris_total}
P_{\mathrm{FARIS}}^{\mathrm{tot}}\triangleq NP_c+N_oP_{\mathrm{DC}}+\xi P_{F,t},\footnote{Note that when $N\leftarrow N_o$, the model reduces to the conventional ARIS power consumption formulation in~\cite{aris5}.}
\end{equation}
where $\xi \triangleq \frac{1}{\upsilon}$ with $\upsilon\in(0,1]$ denoting the amplifier efficiency. Consequently, under the total FARIS power budget $P_{\max,t}$, the reflection design must satisfy $P_{\mathrm{FARIS}}^{\mathrm{tot}} \le P_{\max,t}$, which leads to the equivalent constraint:
\begin{equation}
\label{eq:Pris_constraint}
P_{F,t}\le \upsilon\big(P_{\max,t}-NP_c-N_oP_{\mathrm{DC}}\big)\triangleq P_{\max}.
\end{equation}
Hence, the received signal at $\mathrm{UE}_{t,k}$ is given by
\begin{equation}
\label{rxeq}
\begin{aligned}
y_{t,k}
=&\sqrt{B_{t,k}}\tilde{\mathbf h}_{ru,t,k}^*\mathbf D_{\chi}\mathbf A\mathbf \Phi\tilde{\mathbf h}_{br}x_t\\
&+
\sqrt{B_{ru,t,k}}\tilde{\mathbf h}_{ru,t,k}^*\mathbf D_{\chi}\mathbf A\mathbf \Phi\mathbf n_{F,t}
+\eta_{t,k},
\end{aligned}
\end{equation}
where $\eta_{t,k}\sim\mathcal{CN}(0,\sigma_\eta^2)$ is the receiver noise. Defining the cascaded FARIS-assisted channel gain for $\mathrm{UE}_{t,k}$ by $g_{t,k}\triangleq \tilde{\mathbf h}_{ru,t,k}^*\mathbf D_{\chi}\mathbf A\mathbf \Phi\tilde{\mathbf h}_{br}$, the equivalent received signal becomes
\begin{equation}
\label{ytk}
y_{t,k}
=
\sqrt{B_{t,k}}g_{t,k}\sum_{i\in\mathcal C_t}\sqrt{\rho_{t,i}P_t} x_{t,i}
+z_{F,t,k}+\eta_{t,k},
\end{equation}
where $z_{F,t,k}\triangleq
\sqrt{B_{ru,t,k}}\tilde{\mathbf h}_{ru,t,k}^*\mathbf D_{\chi}\mathbf A\mathbf \Phi\mathbf n_{F,t}$ with variance
\begin{equation}
\label{sftkdef}
\sigma_{F,t,k}^2
=
B_{ru,t,k}\sigma_F^2
\left\|
\tilde{\mathbf h}_{ru,t,k}^*\mathbf D_{\chi}\mathbf A\mathbf \Phi
\right\|_2^2.
\end{equation}
Hence, the total effective noise variance is
\begin{equation}
\label{tenv}
\tilde{\sigma}_{t,k}^2
=
\sigma_\eta^2+
B_{ru,t,k}\sigma_F^2
\left\|
\tilde{\mathbf h}_{ru,t,k}^*\mathbf D_{\chi}\mathbf A\mathbf \Phi
\right\|_2^2.
\end{equation}

Thereafter, within each cluster, the users are re-indexed in descending order of their effective large-scale channel gains~\cite{FRISnoma}: $ B_{t,1}\ge B_{t,2}\ge \cdots \ge B_{t,|\mathcal C_t|}$. 
Under ideal SIC, $\mathrm{UE}_{t,k}$ first decodes and cancels the signals intended for UEs indexed by $i>k$ before decoding its own signal, while the signals intended for UEs indexed by $i<k$ remain as intra-cluster interference. The signal-to-interference-noise (SINR) of $\mathrm{UE}_{t,k}$ is therefore, by~\eqref{ytk} and~\eqref{tenv}, expressed as
\begin{equation}
\label{sinrtk}
\Gamma_{t,k}
=
\frac{
B_{t,k}|g_{t,k}|^2 \rho_{t,k}P_t
}{
B_{t,k}|g_{t,k}|^2 P_t\sum\limits_{i\in\mathcal C_t,i<k}\rho_{t,i}
+\tilde{\sigma}_{t,k}^2}.
\end{equation}
Accordingly, the achievable rate of user $k$ is $R_{t,k}=\log_2(1+\Gamma_{t,k})$, and we ensure the quality-of-service (QoS) by minimum positive rate of $R_{\mathrm{th}}$.

\subsection{Problem Formulation}
The network-wide sum-rate maximization problem for the FARIS-aided downlink NOMA system is formulated as
\begin{equation}
\begin{aligned}
\label{p0}
&\max_{\{\mathcal C_t\},\{\boldsymbol\rho_t\},\boldsymbol{\theta},\boldsymbol{\chi},\boldsymbol{\alpha}}
\mathcal R\triangleq \sum_{t=1}^{C}\sum_{k\in\mathcal C_t}\log_2 \left(1+\Gamma_{t,k}\right) \\
\text{s.t.}~
& \log_2 \left(1+\Gamma_{t,k}\right)\ge R_{\mathrm{th}}~(\forall t,k),~|\mathcal C_t|\le K\\
& \sum_{k\in\mathcal C_t}\rho_{t,k}=1,~ 0< \rho_{t,k}\le 1~(\forall t,k),~\theta_n\in\mathcal Q_b,\\
& 0\le \alpha_n\le \alpha_{\max}\chi^{(n)}~(\forall n),~ \sum_{t=1}^C \xi_t^k =1~(\forall t,k),\\
& \sum_{n=1}^{N}\chi^{(n)}=N_o,~ \chi^{(n)}\in\{0,1\}~(\forall n),~\eqref{eq:Pris_constraint},
\end{aligned}
\end{equation}
where $\mathcal R$ is the network-wide sum-rate,\footnote{Since the clusters are served over orthogonal time slots, let $\tau_t$ denote the transmission duration allocated to cluster $\mathcal C_t$, subject to $\sum_{t=1}^{C}\tau_t\le T_{\max},
~
P_t\tau_t\le E_t~( \forall t)$, where $T_{\max}$ is the maximum transmission interval and $E_t$ is the energy budget of cluster $\mathcal C_t$~\cite{arisnomaiot}. Accordingly, the frame-normalized network sum-rate is $\mathcal R
\triangleq
\frac{1}{T_{\max}}
\sum_{t=1}^{C}\tau_t
\sum_{k\in\mathcal C_t}
\log_2(1+\Gamma_{t,k})$. For fixed physical-layer variables, the cluster spectral efficiencies $\{\mathcal R_t\triangleq\sum_{k\in\mathcal C_t} \log_2(1+\Gamma_{t,k})\}$ are constants, and the optimization with respect to $\{\tau_t\}_{t=1}^{C}$ reduces to a linear program. Since incorporating this standard time-allocation block does not alter the proposed optimization framework, we assume equal-duration time allocation for brevity~\cite{FRISnomaw,FRISnoma}. Specifically, $\tau_t=\frac{T_{\max}}{C}~(\forall t)$ and $\mathcal R$ therefore reduces to the objective of~\eqref{p0}.} $\boldsymbol\rho_t\triangleq[\{\rho_{t,k}\}_{k\in\mathcal C_t}]^{\mathrm T}$, $\boldsymbol\theta = [\theta_1 \cdots  \theta_N]^{\mathrm T}$, $\boldsymbol\alpha=[\alpha_1 \cdots \alpha_N]^{\mathrm{T}}$, and $\boldsymbol\chi=[\chi^{(1)}\cdots \chi^{(N)}]^{\mathrm{T}}$. Problem~\eqref{p0} is an MINLP, where 
the direct solving is prohibitive due to the combinatorial nature of the clustering and selection variables as well as the nonconvex SINR expressions. To address this challenge, we decompose~\eqref{p0} into two stages: NOMA clustering and per-cluster FARIS-NOMA optimization.
\section{Proposed Framework}
\subsection{NOMA Clustering}\label{noc}
To construct NOMA clusters with inherent near-far channel disparity, we exploit the large-scale path-loss characteristics of the FARIS-UE links~\cite{FRISnoma}. Since the large-scale fading is primarily determined by $\{d_{ru,t,k}\}$, it provides a reliable metric for clustering with negligible overhead. Specifically, all UEs are first sorted in ascending order of $\{d_{ru,t,k}\}$, i.e., from the strongest to the weakest large-scale channels. Let $\{u_{(1)},u_{(2)},\cdots,u_{(U)}\}$ denote the ordered UE indices. The clusters are then constructed in an interleaved manner. In particular, the $t$th cluster collects the users with sorted indices
\begin{equation}
\label{sisi}
\mathcal C_t\leftarrow \{u_{(t)},~u_{(t+C)},~u_{(t+2C)},~\cdots\},
\end{equation}
until either the maximum cluster size is reached or all users are assigned. Since the cluster size is limited by $K=\left\lceil \frac{U}{C}\right\rceil$, which guarantees that all UEs can be accommodated within the $C$ orthogonal time slots. This clustering strategy naturally introduces sufficient channel disparity within each cluster, which is beneficial for SIC~\cite{FASnoma11, FRISnoma, FASnoma22}.
\subsection{Per-Cluster FARIS-NOMA Optimization}
After $\{\mathcal C_i\}$ is determined,~\eqref{p0} can be decomposed into independent per-cluster subproblems. For a given $\mathcal C_t$ with its cluster sum-rate $\mathcal R_t \triangleq\sum_{k\in\mathcal C_t}\log_2(1+\Gamma_{t,k})$, the optimization problem reduces to
\begin{equation}
\begin{aligned}
\label{p1}
&\max_{\boldsymbol\rho_t,\boldsymbol\theta,\boldsymbol\chi,\boldsymbol\alpha} \sum_{k\in\mathcal C_t}\log_2(1+\Gamma_{t,k}) \\
\text{s.t.}~
& \log_2 \left(1+\Gamma_{t,k}\right)\ge R_{\mathrm{th}}~(\forall t,k),\\
& \sum_{k\in\mathcal C_t}\rho_{t,k}=1,~ 0< \rho_{t,k}\le 1~(\forall t,k),\\
& \theta_n\in\mathcal Q_b,~ 0\le \alpha_n\le \alpha_{\max}\chi^{(n)}~(\forall n),\\
& \sum_{n=1}^{N}\chi^{(n)}=N_o,~ \chi^{(n)}\in\{0,1\}~(\forall n),~\eqref{eq:Pris_constraint}.\\
\end{aligned}
\end{equation}
Since~\eqref{p1} still remains highly nonconvex due to the coupled optimization variables, we adopt an AO framework that sequentially updates the variables.
\subsubsection{Update of $\boldsymbol\rho_t$}\label{secrho}
For fixed $(\boldsymbol\theta,\boldsymbol\chi,\boldsymbol\alpha)$, $\{g_{t,k}\}_{k\in\mathcal C_t}$ and $\{\tilde\sigma_{t,k}^2\}_{k\in\mathcal C_t}$ are fixed. Defining $a_{t,k}\triangleq B_{t,k}|g_{t,k}|^2P_t,~
c_{t,k}\triangleq \frac{\tilde\sigma_{t,k}^2}{a_{t,k}}$, the SINR in~\eqref{sinrtk} can be rewritten as
\begin{equation}
\label{sinr_rho}
\Gamma_{t,k}
=
\frac{\rho_{t,k}}
{\sum_{i\in\mathcal C_t,i<k}\rho_{t,i}+c_{t,k}}.
\end{equation}
Accordingly, $R_{t,k}$ becomes $R_{t,k}
=
\log_2\left(
1+
\frac{\rho_{t,k}}
{\sum_{i\in\mathcal C_t,i<k}\rho_{t,i}+c_{t,k}}
\right)$, and the power-allocation subproblem is thus expressed as
\begin{equation}
\begin{aligned}
\label{prho2}
&\max_{\boldsymbol\rho_t} \sum_{k\in\mathcal C_t}\log_2(1+\Gamma_{t,k})\\
\text{s.t.}~
&\sum_{k\in\mathcal C_t}\rho_{t,k}=1,~\rho_{t,k}\ge 0,~\Gamma_{t,k}\ge \gamma_{\mathrm{th}}~ (\forall k\in\mathcal C_t),
\end{aligned}
\end{equation}
where $\gamma_{\mathrm{th}}\triangleq 2^{R_{\mathrm{th}}}-1$. To efficiently solve~\eqref{prho2}, we resort to GP, which is well-suited for handling posynomial objective and constraint structures after appropriate transformations~\cite{gp, iotnoma}. Since $\log_2(\cdot)$ is monotonically increasing, maximizing the objective of~\eqref{prho2} is equivalent to maximizing the product of $(1+\Gamma_{t,k})$, i.e., $\max_{\boldsymbol\rho_t} \prod_{k\in\mathcal C_t}(1+\Gamma_{t,k})$. Using~\eqref{sinr_rho}, we have
\begin{equation}
\label{one_plus_gamma}
1+\Gamma_{t,k}
=
\frac{\sum_{i\in\mathcal C_t,i\le k}\rho_{t,i}+c_{t,k}}{\sum_{i\in\mathcal C_t,i<k}\rho_{t,i}+c_{t,k}}.
\end{equation}
To cast the problem into a GP-compatible form, we introduce auxiliary variables $\{\tau_{t,k}(>0)\}_{k\in\mathcal C_t}$ satisfying $\tau_{t,k}\le 1+\Gamma_{t,k}~ (\forall k\in\mathcal C_t)$. Then~\eqref{prho2} is equivalently transformed into
\begin{equation}
\begin{aligned}
\label{prho_aux}
\max_{\boldsymbol\rho_t,\boldsymbol\tau_t}
&\prod_{k\in\mathcal C_t}\tau_{t,k}\\
\text{s.t.}~
&\tau_{t,k}\le
\frac{\sum_{i\in\mathcal C_t,i\le k}\rho_{t,i}+c_{t,k}}
{\sum_{i\in\mathcal C_t,i<k}\rho_{t,i}+c_{t,k}}~(\forall k\in\mathcal C_t),\\
&\frac{\gamma_{\mathrm{th}}\left(\sum_{i\in\mathcal C_t,i<k}\rho_{t,i}+c_{t,k}\right)}
{\rho_{t,k}}\le 1~(\forall k\in\mathcal C_t),\\
&\sum_{k\in\mathcal C_t}\rho_{t,k}=1,~ \rho_{t,k}\ge 0~ (\forall k\in\mathcal C_t).
\end{aligned}
\end{equation}
Note that the QoS constraints have already been converted into standard posynomial inequalities.

The remaining difficulty lies in the fractional constraint involving $\tau_{t,k}$. To handle this, define $u_{t,k}(\boldsymbol\rho_t)\triangleq\sum_{i\in\mathcal C_t,i\le k}\rho_{t,i}+c_{t,k}$. Since $u_{t,k}(\boldsymbol\rho_t)$ is a posynomial, it can be lower-bounded by a monomial via the arithmetic-geometric mean approximation (AGMA). Specifically, at the $r$th GP iteration with a feasible point $\boldsymbol\rho_t^{(r)}$, we construct
\begin{equation}
\label{uk_hat}
u_{t,k}(\boldsymbol\rho_t)
\ge
\hat u_{t,k}\big(\boldsymbol\rho_t;\boldsymbol\rho_t^{(r)}\big),
\end{equation}
where $\hat u_{t,k}\big(\boldsymbol\rho_t;\boldsymbol\rho_t^{(r)}\big)
=
\left(\frac{c_{t,k}}{\beta_{t,k,0}^{(r)}}\right)^{\beta_{t,k,0}^{(r)}}
\prod_{i\in\mathcal C_t,i\le k}
\left(\frac{\rho_{t,i}}{\beta_{t,k,i}^{(r)}}\right)^{\beta_{t,k,i}^{(r)}}$, with weights $\beta_{t,k,0}^{(r)}
=
\frac{c_{t,k}}{u_{t,k}(\boldsymbol\rho_t^{(r)})},
~
\beta_{t,k,i}^{(r)}
=
\frac{\rho_{t,i}^{(r)}}{u_{t,k}(\boldsymbol\rho_t^{(r)})}~( i\in\mathcal C_t, i\le k)$, which satisfy $\beta_{t,k,0}^{(r)}+\sum_{i\in\mathcal C_t,i\le k}\beta_{t,k,i}^{(r)}=1$. Then, the constraint $\tau_{t,k}\le
\frac{u_{t,k}(\boldsymbol\rho_t)}
{\sum_{i\in\mathcal C_t,i<k}\rho_{t,i}+c_{t,k}}$ in~\eqref{prho_aux} is conservatively approximated by
\begin{equation}
\label{tau_gp_cons}
\frac{
\tau_{t,k}\left(\sum_{i\in\mathcal C_t,i<k}\rho_{t,i}+c_{t,k}\right)
}{
\hat u_{t,k}\big(\boldsymbol\rho_t;\boldsymbol\rho_t^{(r)}\big)
}
\le 1~(\forall k\in\mathcal C_t),
\end{equation}
which is a standard GP constraint since the numerator is a posynomial and the denominator is a monomial. Moreover, since the objective is increasing in $\{\rho_{t,k}\}$, the total-power constraint is active at optimum, and the equality $\sum_{k\in\mathcal C_t}\rho_{t,k}=1$ can be replaced by $\sum_{k\in\mathcal C_t}\rho_{t,k}\le 1$ without loss of optimality, yielding a standard GP form.

Therefore, at iteration $r$, the power-allocation update is obtained by solving the following GP:
\begin{equation}
\begin{aligned}
\label{gp_rho}
\min_{\boldsymbol\rho_t,\boldsymbol\tau_t}
&\prod_{k\in\mathcal C_t}\tau_{t,k}^{-1}\\
\text{s.t.}~
&\frac{
\tau_{t,k}\left(\sum_{i\in\mathcal C_t,i<k}\rho_{t,i}+c_{t,k}\right)
}{
\hat u_{t,k}\big(\boldsymbol\rho_t;\boldsymbol\rho_t^{(r)}\big)
}
\le 1~(\forall k\in\mathcal C_t),\\
&\frac{\gamma_{\mathrm{th}}\left(\sum_{i\in\mathcal C_t,i<k}\rho_{t,i}+c_{t,k}\right)}
{\rho_{t,k}}\le 1~(\forall k\in\mathcal C_t),\\
&\sum_{k\in\mathcal C_t}\rho_{t,k}\le 1,\rho_{t,k}> 0, \tau_{t,k}>0~(\forall k\in\mathcal C_t).
\end{aligned}
\end{equation}
By introducing the logarithmic change of variables $\bar{\rho}_{t,k}=\log \rho_{t,k},~\bar{\tau}_{t,k}=\log \tau_{t,k}$, the GP can be equivalently transformed into a convex problem in the log-domain by taking log of the objective and constraint functions, where each posynomial constraint becomes a convex log-sum-exp function~\cite{gp}. The resulting convex problem can therefore be efficiently solved using interior-point methods~\cite{boyd}. After solving~\eqref{gp_rho}, the approximation point is updated as $\boldsymbol\rho_t^{(r+1)}\leftarrow \boldsymbol\rho_t^\star$, where $\boldsymbol\rho_t^\star$ is the solution of~\eqref{gp_rho}. The above procedure is repeated until convergence. Since the AGMA-based monomial approximation is tight at the current $\boldsymbol\rho^{(r)}$ by definition, and combining with~\eqref{uk_hat}, the GP procedure yields a nondecreasing objective sequence and converges to a stationary point of~\eqref{prho2}.

\subsubsection{Update of $\boldsymbol\alpha$}\label{sasa}
For fixed $(\boldsymbol\theta,\boldsymbol\chi,\boldsymbol\rho_t)$, $\boldsymbol\alpha$ affects both the desired signal power $B_{t,k}|g_{t,k}|^2\rho_{t,k}P_t$ and the forwarded amplifier-noise power
$\sigma_{F,t,k}^2=B_{ru,t,k}\sigma_F^2
\left\|
\tilde{\mathbf h}_{ru,t,k}^*\mathbf D_\chi\mathbf A\mathbf\Phi
\right\|_2^2$.
Hence, the amplification-gain optimization remains nonconvex. To make the dependence on $\boldsymbol\alpha$ explicit, define the effective cascaded channel vector without amplification for $\mathrm{UE}_{t,k}$ as
\begin{equation}
\label{ctk}
\mathbf c_{t,k}
\triangleq
\mathbf D_{\chi}\mathbf\Phi
\big(\tilde{\mathbf h}_{br}\odot\tilde{\mathbf h}_{ru,t,k}^{*}\big),
\end{equation}
where $\odot$ denotes the Hadamard product. Then, $g_{t,k}$ can be rewritten as $g_{t,k}
= \boldsymbol\alpha^{\mathrm T}\mathbf c_{t,k}$. Accordingly, the desired signal power term in~\eqref{sinrtk} becomes
\begin{equation}
\label{sig_alpha}
B_{t,k}P_t\rho_{t,k}|g_{t,k}|^2
=
\boldsymbol\alpha^*\mathbf S_{t,k}\boldsymbol\alpha,
\end{equation}
where
$\mathbf S_{t,k}\triangleq
B_{t,k}P_t\rho_{t,k}\mathbf c_{t,k}\mathbf c_{t,k}^*\succeq\mathbf 0$.
Likewise, the intra-cluster interference term can be expressed as
\begin{equation}
\label{int_alpha}
B_{t,k}P_t
\sum_{\substack{i\in\mathcal C_t\\i<k}}
\rho_{t,i}|g_{t,k}|^2
=
\boldsymbol\alpha^*\mathbf I_{t,k}\boldsymbol\alpha,
\end{equation}
where
$\mathbf I_{t,k}\triangleq
B_{t,k}P_t
\big(\sum_{i\in\mathcal C_t,i<k}\rho_{t,i}\big)
\mathbf c_{t,k}\mathbf c_{t,k}^*\succeq\mathbf 0$.
Moreover, from~\eqref{sftkdef} and~\eqref{tenv}, the forwarded amplifier-noise power is
\begin{equation}
\label{noise_alpha}
B_{ru,t,k}\sigma_F^2
\left\|
\tilde{\mathbf h}_{ru,t,k}^{*}
\mathbf D_{\chi}\mathbf A\mathbf\Phi
\right\|_2^2
=
\boldsymbol\alpha^*\mathbf Z_{t,k}\boldsymbol\alpha,
\end{equation}
where
\begin{equation}
\label{Ztk}
\begin{aligned}
\mathbf Z_{t,k}\triangleq B_{ru,t,k}\sigma_F^2\operatorname{diag}\Big(&
\chi^{(1)}|[\tilde{\mathbf h}_{ru,t,k}]_1|^2,\\
&\cdots,
\chi^{(N)}|[\tilde{\mathbf h}_{ru,t,k}]_N|^2
\Big)\succeq\mathbf 0.
\end{aligned}
\end{equation}
Therefore, the SINR of $\mathrm{UE}_{t,k}$ can be compactly expressed as
\begin{equation}
\label{sinr_alpha}
\Gamma_{t,k}(\boldsymbol\alpha)
=
\frac{\boldsymbol\alpha^*\mathbf S_{t,k}\boldsymbol\alpha}
{\boldsymbol\alpha^*\mathbf N_{t,k}\boldsymbol\alpha+\sigma_\eta^2},
\end{equation}
where $\mathbf N_{t,k}\triangleq\mathbf I_{t,k}+\mathbf Z_{t,k}\succeq\mathbf 0$.
Accordingly, the amplification-gain optimization subproblem is given by
\begin{equation}
\begin{aligned}
\label{pa0}
\max_{\boldsymbol\alpha}~
&\sum_{k\in\mathcal C_t}
\log_2\big(1+\Gamma_{t,k}(\boldsymbol\alpha)\big)\\
\mathrm{s.t.}~
&0\le\alpha_n\le\alpha_{\max}\chi^{(n)}~ (\forall n),~\boldsymbol\alpha^{\mathrm T}\mathbf C_t\boldsymbol\alpha\le P_{\max},\\
&\Gamma_{t,k}(\boldsymbol\alpha)\ge\gamma_{\mathrm{th}}~ (\forall k\in\mathcal C_t),
\end{aligned}
\end{equation}
where $\gamma_{\mathrm{th}}\triangleq2^{R_{\mathrm{th}}}-1$, and
\begin{equation}
\label{Ct}
\begin{aligned}
\mathbf C_t
\triangleq
\operatorname{diag}\Big(&
\chi^{(1)}
\big(B_{br}P_t|[\tilde{\mathbf h}_{br}]_1|^2+\sigma_F^2\big),\\
&\cdots,
\chi^{(N)}
\big(B_{br}P_t|[\tilde{\mathbf h}_{br}]_N|^2+\sigma_F^2\big)
\Big)\succeq\mathbf 0
\end{aligned}
\end{equation}
is the FARIS power-constraint matrix. Note that by~\eqref{sinr_alpha}, the QoS constraint in~\eqref{pa0} is equivalently written as
\begin{equation}
\label{qos_alpha_exact}
\boldsymbol\alpha^*\mathbf S_{t,k}\boldsymbol\alpha
\ge
\gamma_{\mathrm{th}}
\left(
\boldsymbol\alpha^*\mathbf N_{t,k}\boldsymbol\alpha
+\sigma_\eta^2
\right)~( \forall k\in\mathcal C_t).
\end{equation}

Problem~\eqref{pa0} is still nonconvex due to the sum of logarithms of fractional quadratic functions and the nonconvex QoS constraints in~\eqref{qos_alpha_exact}. To handle the objective function, we first apply the Lagrangian dual transform used in fractional programming (FP)~\cite{fracp}. Introducing auxiliary variables $\{\lambda_{t,k}\ge0\}_{k\in\mathcal C_t}$, we equivalently rewrite~\eqref{pa0} as
\begin{equation}
\begin{aligned}
\label{pa1}
&\max_{\boldsymbol\alpha,\boldsymbol\lambda_t}
\sum_{k\in\mathcal C_t}
\Bigg[
\log_2(1+\lambda_{t,k})
-\frac{\lambda_{t,k}}{\ln 2}\\
&~~~~~~~~~~~~~~+\frac{1+\lambda_{t,k}}{\ln 2}
\frac{\boldsymbol\alpha^*\mathbf S_{t,k}\boldsymbol\alpha}
{\boldsymbol\alpha^*(\mathbf S_{t,k}+\mathbf N_{t,k})\boldsymbol\alpha
+\sigma_\eta^2}
\Bigg]\\
\mathrm{s.t.}~
&0\le\alpha_n\le\alpha_{\max}\chi^{(n)}~ (\forall n),~\boldsymbol\alpha^{\mathrm T}\mathbf C_t\boldsymbol\alpha\le P_{\max},\\
&\Gamma_{t,k}(\boldsymbol\alpha)\ge\gamma_{\mathrm{th}}~ (\forall k\in\mathcal C_t),
\end{aligned}
\end{equation}
where the optimal auxiliary variable is given by $\lambda_{t,k}^{\star}
=
\Gamma_{t,k}(\boldsymbol\alpha)$. Although~\eqref{pa1} removes the logarithmic dependence on the SINR ratios, fractional quadratic terms still remain. We therefore apply the quadratic transform~\cite{fracp}. Define $\mathbf s_{t,k}
\triangleq
\sqrt{B_{t,k}P_t\rho_{t,k}} \mathbf c_{t,k}$ and $\mathbf T_{t,k}
\triangleq
\mathbf S_{t,k}+\mathbf N_{t,k}\succeq\mathbf 0$, where $\mathbf S_{t,k}=\mathbf s_{t,k}\mathbf s_{t,k}^*$.
Introducing a second set of auxiliary variables
$\{y_{t,k}\in\mathbb C\}_{k\in\mathcal C_t}$, the fractional term can be transformed as
\begin{equation}
\label{qt_identity}
\begin{aligned}
&\frac{(1+\lambda_{t,k})
\boldsymbol\alpha^*\mathbf S_{t,k}\boldsymbol\alpha}
{\boldsymbol\alpha^*\mathbf T_{t,k}\boldsymbol\alpha+\sigma_\eta^2}\\
&=\max_{y_{t,k}\in\mathbb C}
\Bigg[
2\sqrt{1+\lambda_{t,k}}
\Re\big\{y_{t,k}^{*}\mathbf s_{t,k}^*\boldsymbol\alpha\big\}
-|y_{t,k}|^2
\big(
\boldsymbol\alpha^*\mathbf T_{t,k}\boldsymbol\alpha
+\sigma_\eta^2
\big)
\Bigg].
\end{aligned}
\end{equation}
The optimal $y_{t,k}$ is obtained in closed-form as
\begin{equation}
\label{y_opt}
y_{t,k}^{\star}
=
\frac{
\sqrt{1+\lambda_{t,k}} 
\mathbf s_{t,k}^*\boldsymbol\alpha
}
{
\boldsymbol\alpha^*\mathbf T_{t,k}\boldsymbol\alpha+\sigma_\eta^2
}.
\end{equation}
Substituting~\eqref{qt_identity} into~\eqref{pa1}, the problem is equivalently transformed into
\begin{equation}
\begin{aligned}
\label{pa2}
&\max_{\boldsymbol\alpha,\boldsymbol\lambda_t,\mathbf y_t}
\sum_{k\in\mathcal C_t}
\Bigg[
\log_2(1+\lambda_{t,k})
-\frac{\lambda_{t,k}}{\ln 2}\\
&~~~~~~~~~~~~~~~~+\frac{1}{\ln 2}
\Big(
2\sqrt{1+\lambda_{t,k}}
\Re\{y_{t,k}^{*}\mathbf s_{t,k}^*\boldsymbol\alpha\}\\
&~~~~~~~~~~~~~~~~-|y_{t,k}|^2
(\boldsymbol\alpha^*\mathbf T_{t,k}\boldsymbol\alpha+\sigma_\eta^2)
\Big)
\Bigg]\\
\mathrm{s.t.}~
&0\le\alpha_n\le\alpha_{\max}\chi^{(n)}~ (\forall n),~\boldsymbol\alpha^{\mathrm T}\mathbf C_t\boldsymbol\alpha\le P_{\max},\\
&\lambda_{t,k}\ge0~ (\forall k\in\mathcal C_t),~\Gamma_{t,k}(\boldsymbol\alpha)\ge\gamma_{\mathrm{th}}~ (\forall k\in\mathcal C_t).
\end{aligned}
\end{equation}
For fixed $(\boldsymbol\lambda_t,\mathbf y_t)$, the objective function in~\eqref{pa2} is a concave quadratic function of $\boldsymbol\alpha$. In particular, the term
$\sum_{k\in\mathcal C_t}2\sqrt{1+\lambda_{t,k}}
\Re\{y_{t,k}^{*}\mathbf s_{t,k}^*\boldsymbol\alpha\}$
is affine in $\boldsymbol\alpha$, whereas
$-\sum_{k\in\mathcal C_t}|y_{t,k}|^2
\boldsymbol\alpha^*\mathbf T_{t,k}\boldsymbol\alpha$
is concave because $\mathbf T_{t,k}\succeq\mathbf 0$.

It remains to handle the nonconvex QoS constraints. Since
$\mathbf S_{t,k}\succeq\mathbf 0$, $\boldsymbol\alpha^*\mathbf S_{t,k}\boldsymbol\alpha$
is convex in $\boldsymbol\alpha$ and admits the following global affine lower-bound at a feasible point $\boldsymbol\alpha^{(q)}$:
\begin{equation}
\label{signal_alpha_lb}
\begin{aligned}
\boldsymbol\alpha^*\mathbf S_{t,k}\boldsymbol\alpha
\ge
\underline{s}_{t,k}^{(q)}(\boldsymbol\alpha)
\triangleq{}&
2\Re\Big\{
(\boldsymbol\alpha^{(q)})^*
\mathbf S_{t,k}\boldsymbol\alpha
\Big\}-(\boldsymbol\alpha^{(q)})^*
\mathbf S_{t,k}\boldsymbol\alpha^{(q)}.
\end{aligned}
\end{equation}
Consequently,~\eqref{qos_alpha_exact} is conservatively approximated by
\begin{equation}
\label{qos_alpha_sca}
\gamma_{\mathrm{th}}
\left(
\boldsymbol\alpha^*\mathbf N_{t,k}\boldsymbol\alpha
+\sigma_\eta^2
\right)
\le
\underline{s}_{t,k}^{(q)}(\boldsymbol\alpha),
~ \forall k\in\mathcal C_t.
\end{equation}
The left-hand side of~\eqref{qos_alpha_sca} is convex and its right-hand side is affine in $\boldsymbol\alpha$; hence,~\eqref{qos_alpha_sca} is a convex quadratic constraint. Moreover, the approximation is tight at $\boldsymbol\alpha=\boldsymbol\alpha^{(q)}$. Therefore, if $\boldsymbol\alpha^{(q)}$ satisfies the original QoS constraints, it also satisfies~\eqref{qos_alpha_sca}, and every point satisfying~\eqref{qos_alpha_sca} is feasible for~\eqref{qos_alpha_exact}.

For fixed $(\boldsymbol\lambda_t,\mathbf y_t)$ and the approximation point $\boldsymbol\alpha^{(q)}$, the update of $\boldsymbol\alpha$ thus reduces to the following convex quadratically constrained quadratic program (QCQP):
\begin{equation}
\begin{aligned}
\label{pa3}
\max_{\boldsymbol\alpha}~
&2\Re\{\mathbf q_t^*\boldsymbol\alpha\}
-\boldsymbol\alpha^*\mathbf M_t\boldsymbol\alpha\\
\mathrm{s.t.}~
&0\le\alpha_n\le\alpha_{\max}\chi^{(n)}~ (\forall n),~\boldsymbol\alpha^{\mathrm T}\mathbf C_t\boldsymbol\alpha\le P_{\max},\\
&\gamma_{\mathrm{th}}
\left(
\boldsymbol\alpha^*\mathbf N_{t,k}\boldsymbol\alpha
+\sigma_\eta^2
\right)
\le\underline{s}_{t,k}^{(q)}(\boldsymbol\alpha)~( \forall k\in\mathcal C_t),
\end{aligned}
\end{equation}
where $\mathbf q_t
\triangleq
\frac{1}{\ln 2}
\sum_{k\in\mathcal C_t}
\sqrt{1+\lambda_{t,k}} y_{t,k}\mathbf s_{t,k}$ and $\mathbf M_t
\triangleq
\frac{1}{\ln 2}
\sum_{k\in\mathcal C_t}
|y_{t,k}|^2\mathbf T_{t,k}\succeq\mathbf 0$. Therefore, the amplification-gain update is carried out iteratively as follows. Starting from a feasible $\boldsymbol\alpha^{(0)}$, for a given $\boldsymbol\alpha^{(q)}$, the auxiliary variables are updated according to
\begin{equation}
\label{lambda_update}
\lambda_{t,k}^{(q+1)}
=
\Gamma_{t,k}\big(\boldsymbol\alpha^{(q)}\big)~ (\forall k\in\mathcal C_t),
\end{equation}
and
\begin{equation}
\label{y_update}
y_{t,k}^{(q+1)}
=
\frac{
\sqrt{1+\lambda_{t,k}^{(q+1)}} 
\mathbf s_{t,k}^*\boldsymbol\alpha^{(q)}
}
{
(\boldsymbol\alpha^{(q)})^*
\mathbf T_{t,k}\boldsymbol\alpha^{(q)}
+\sigma_\eta^2
}~ (\forall k\in\mathcal C_t).
\end{equation}
Then, for fixed $\boldsymbol\lambda_t^{(q+1)}$ and $\mathbf y_t^{(q+1)}$, the convex QCQP~\eqref{pa3}, whose QoS constraints are constructed at $\boldsymbol\alpha^{(q)}$, is solved to obtain $\boldsymbol\alpha^{(q+1)}\leftarrow\boldsymbol\alpha^\star$, where $\boldsymbol\alpha^\star$ denotes the solution of~\eqref{pa3}. The above steps are repeated until convergence. Since every transform is tight at the current iterate~\cite{fracp}, each update yields a nondecreasing objective value. Hence, the resulting iterative procedure generates a nondecreasing sum-rate sequence and converges to a stationary point of~\eqref{pa0}.

\subsubsection{Update of $\boldsymbol\theta$}
For fixed $(\boldsymbol\alpha,\boldsymbol\chi,\boldsymbol\rho_t)$, $\boldsymbol\theta$ determines $g_{t,k}$, which can be rewritten as
\begin{equation}
\label{g_phase}
g_{t,k}
=
\sum_{n=1}^{N}
\chi^{(n)}\alpha_n
\tilde h_{ru,t,k,n}^{*}\tilde h_{br,n}
e^{j\theta_n}.
\end{equation}
Define
$\boldsymbol\phi(\boldsymbol\theta)
\triangleq[e^{j\theta_1} \cdots e^{j\theta_N}]^{\mathrm T}$. Then,~\eqref{g_phase} becomes $g_{t,k}=\mathbf d_{t,k}^*\boldsymbol\phi$, where
\begin{equation}
\label{dtk}
\mathbf d_{t,k}
\triangleq\left[\chi^{(1)}\alpha_1
\tilde h_{ru,t,k,1}\tilde h_{br,1}^{*}\cdots\chi^{(N)}\alpha_N
\tilde h_{ru,t,k,N}\tilde h_{br,N}^{*}
\right]^{\rm T}.
\end{equation}
Accordingly, $|g_{t,k}|^2=
\boldsymbol\phi^*\mathbf Q_{t,k}\boldsymbol\phi$ and $
\mathbf Q_{t,k}
\triangleq
\mathbf d_{t,k}\mathbf d_{t,k}^*
\succeq\mathbf 0$. Now let $s_{t,k}\triangleq B_{t,k}P_t\rho_{t,k}$ and $i_{t,k}\triangleq
B_{t,k}P_t
\sum_{\substack{j\in\mathcal C_t\\j<k}}\rho_{t,j}$. For fixed $(\boldsymbol\alpha,\boldsymbol\chi)$, $\tilde\sigma_{t,k}^2$ in~\eqref{tenv} is independent of $\boldsymbol\theta$, because the forwarded amplifier-noise power in~\eqref{sftkdef} depends only on the magnitudes of the entries of $\boldsymbol\Phi$. Hence, defining $\nu_{t,k}\triangleq\tilde\sigma_{t,k}^2$, the SINR in~\eqref{sinrtk} can be expressed as
\begin{equation}
\label{sinr_phi}
\Gamma_{t,k}(\boldsymbol\phi)
=
\frac{
s_{t,k}\boldsymbol\phi^*\mathbf Q_{t,k}\boldsymbol\phi
}{
i_{t,k}\boldsymbol\phi^*\mathbf Q_{t,k}\boldsymbol\phi
+\nu_{t,k}
}.
\end{equation}

Now define $\mathcal F_b
\triangleq
\{e^{j\varphi}:\varphi\in\mathcal Q_b\}$. The QoS-constrained phase optimization problem is then formulated as
\begin{equation}
\label{p_phase}
\begin{aligned}
\max_{\boldsymbol\phi}~
&f(\boldsymbol\phi)
\triangleq
\sum_{k\in\mathcal C_t}
\log_2\big(1+\Gamma_{t,k}(\boldsymbol\phi)\big)\\
\mathrm{s.t.}~
&\Gamma_{t,k}(\boldsymbol\phi)
\ge\gamma_{\mathrm{th}}~( \forall k\in\mathcal C_t),~\phi_n\in\mathcal F_b~( \forall n),
\end{aligned}
\end{equation}
where $\gamma_{\mathrm{th}}\triangleq2^{R_{\mathrm{th}}}-1$.
Due to $\mathcal Q_b$,~\eqref{p_phase} is combinatorial. We hence first relax $\mathcal F_b$ to the continuous unit-modulus set
$\{\phi_n\in\mathbb C:|\phi_n|=1,~\forall n\}$
while retaining all QoS constraints. Using~\eqref{sinr_phi}, the objective function can be rewritten as
\begin{equation}
\label{fphi_ratio}
f(\boldsymbol\phi)
=
\sum_{k\in\mathcal C_t}
\log_2
\left(
\frac{
(s_{t,k}+i_{t,k})
\boldsymbol\phi^*\mathbf Q_{t,k}\boldsymbol\phi
+\nu_{t,k}
}{
i_{t,k}\boldsymbol\phi^*\mathbf Q_{t,k}\boldsymbol\phi
+\nu_{t,k}
}
\right).
\end{equation}
The gradient of $f(\boldsymbol\phi)$ with respect to $\boldsymbol\phi^*$ is given in~\eqref{grad_phi}.
\begin{figure*}
\begin{equation}
\label{grad_phi}
\nabla_{\boldsymbol\phi^*}f(\boldsymbol\phi)
=
\frac{1}{\ln 2}
\sum_{k\in\mathcal C_t}
\frac{
s_{t,k}\nu_{t,k}
}{
\big((s_{t,k}+i_{t,k})
\boldsymbol\phi^*\mathbf Q_{t,k}\boldsymbol\phi
+\nu_{t,k}\big)
\big(i_{t,k}
\boldsymbol\phi^*\mathbf Q_{t,k}\boldsymbol\phi
+\nu_{t,k}\big)
}
\mathbf Q_{t,k}\boldsymbol\phi.
\end{equation}
\hrule
\end{figure*}

We next express the QoS constraints suitable for inner approximation. From~\eqref{sinr_phi}, it is equivalent to
\begin{equation}
\label{qos_phi_exact0}
\big(s_{t,k}-\gamma_{\mathrm{th}}i_{t,k}\big)
\boldsymbol\phi^*\mathbf Q_{t,k}\boldsymbol\phi
\ge
\gamma_{\mathrm{th}}\nu_{t,k}.
\end{equation}
Since $\nu_{t,k}>0$, a necessary condition for feasibility is
\begin{equation}
\label{delta_positive}
\delta_{t,k}
\triangleq
s_{t,k}-\gamma_{\mathrm{th}}i_{t,k}>0~ (\forall k\in\mathcal C_t).
\end{equation}
Condition~\eqref{delta_positive} is automatically satisfied when the phase-update block is initialized at a point feasible for the original QoS constraints. Under~\eqref{delta_positive}, define $\xi_{t,k}
\triangleq
\frac{\gamma_{\mathrm{th}}\nu_{t,k}}{\delta_{t,k}}$. Then,~\eqref{qos_phi_exact0} reduces to
\begin{equation}
\label{qos_phi_exact}
h_{t,k}(\boldsymbol\phi)\triangleq\boldsymbol\phi^*\mathbf Q_{t,k}\boldsymbol\phi
\ge\xi_{t,k}~( \forall k\in\mathcal C_t).
\end{equation}
Since $\mathbf Q_{t,k}\succeq\mathbf 0$, $h_{t,k}(\boldsymbol\phi)$ is a convex quadratic function of $\boldsymbol\phi$ and admits the affine lower-bound at the current iterate $\boldsymbol\phi^{(m)}$:
\begin{equation}
\label{h_phi_lb}
\begin{aligned}
&h_{t,k}(\boldsymbol\phi)\\
&\ge\underline h_{t,k}
(\boldsymbol\phi|\boldsymbol\phi^{(m)})
\triangleq
2\Re\left\{
(\boldsymbol\phi^{(m)})^*
\mathbf Q_{t,k}\boldsymbol\phi
\right\}-(\boldsymbol\phi^{(m)})^*
\mathbf Q_{t,k}\boldsymbol\phi^{(m)}.
\end{aligned}
\end{equation}
Therefore, the inner constraint:
\begin{equation}
\label{qos_phi_inner}
\underline h_{t,k}
(\boldsymbol\phi|\boldsymbol\phi^{(m)})
\ge\xi_{t,k}~( \forall k\in\mathcal C_t),
\end{equation}
is sufficient for the original QoS constraint in~\eqref{qos_phi_exact}.

Following the MM framework based on quadratic minorization~\cite{mmtsp,mmtit}, let $\beta^{(m)}>0$ be any constant no smaller than a Lipschitz constant of $\nabla_{\boldsymbol\phi^*} f(\boldsymbol\phi)$ at $\boldsymbol\phi^{(m)}$. Then, a valid minorizing surrogate of $f(\boldsymbol\phi)$ at $\boldsymbol\phi^{(m)}$ is given by
\begin{equation}
\label{surrogate_phi}
\begin{aligned}
\widetilde f
(\boldsymbol\phi|\boldsymbol\phi^{(m)})
\triangleq&
f(\boldsymbol\phi^{(m)})+2\Re\left\{
\left(
\nabla_{\boldsymbol\phi^*}f(\boldsymbol\phi^{(m)})
\right)^*
(\boldsymbol\phi-\boldsymbol\phi^{(m)})
\right\}\\
&-\beta^{(m)}
\|\boldsymbol\phi-\boldsymbol\phi^{(m)}\|_2^2.
\end{aligned}
\end{equation}
Hence, maximizing~\eqref{surrogate_phi} subject to the corresponding constraints is equivalent to
\begin{equation}
\label{mm_prob}
\begin{aligned}
\max_{\boldsymbol\phi}~
&\Re\left\{\boldsymbol\phi^*\mathbf v^{(m)}\right\}\\
\mathrm{s.t.}~
&\underline h_{t,k}
(\boldsymbol\phi|\boldsymbol\phi^{(m)})
\ge\xi_{t,k}~( \forall k\in\mathcal C_t),~|\phi_n|=1~(\forall n),
\end{aligned}
\end{equation}
where $\mathbf v^{(m)}
\triangleq
\nabla_{\boldsymbol\phi^*}f(\boldsymbol\phi^{(m)})
+\beta^{(m)}\boldsymbol\phi^{(m)}$. To solve~\eqref{mm_prob}, let $L\triangleq|\mathcal Q_b|=2^b$ and enumerate the available phase angles as
$\mathcal Q_b=\{\varphi_1,\cdots,\varphi_L\}$ with $u_l\triangleq e^{j\varphi_l}$. Introduce the one-hot binary variables
$z_{n,l}\in\{0,1\}$ such that
\begin{equation}
\label{onehot_phase}
\phi_n
=
\sum_{l=1}^{L}u_l z_{n,l},
~
\sum_{l=1}^{L}z_{n,l}=1~ (\forall n).
\end{equation}
This enforces $\phi_n\in\mathcal F_b$ exactly. We further define
\begin{equation}
\label{milp_coeff_obj}
c_{n,l}^{(m)}
\triangleq
\Re\left\{u_l^*v_n^{(m)}\right\},~d_{t,k,n,l}^{(m)}
\triangleq
2\Re\left\{
\left[
\mathbf Q_{t,k}\boldsymbol\phi^{(m)}
\right]_n^*u_l
\right\}.
\end{equation}
Using~\eqref{onehot_phase}, the MM objective in~\eqref{mm_prob} becomes
$\sum_{n=1}^{N}\sum_{l=1}^{L}c_{n,l}^{(m)}z_{n,l}$.
Moreover,
~\eqref{qos_phi_inner} is equivalently:
\begin{equation}
\label{milp_qos_constraint}
\sum_{n=1}^{N}\sum_{l=1}^{L}
d_{t,k,n,l}^{(m)}z_{n,l}
\ge
\xi_{t,k}+h_{t,k}(\boldsymbol\phi^{(m)})~ (\forall k\in\mathcal C_t).
\end{equation}
Consequently, the joint discrete phase update at the $m$th MM iteration is obtained by solving the following MILP:
\begin{equation}
\label{mm_phase_milp}
\begin{aligned}
\max_{\{z_{n,l}\}}~
&\sum_{n=1}^{N}\sum_{l=1}^{L}
c_{n,l}^{(m)}z_{n,l}\\
\mathrm{s.t.}~
&\sum_{n=1}^{N}\sum_{l=1}^{L}
d_{t,k,n,l}^{(m)}z_{n,l}
\ge
\xi_{t,k}+h_{t,k}(\boldsymbol\phi^{(m)})~(\forall k\in\mathcal C_t),\\
&\sum_{l=1}^{L}z_{n,l}=1~ (\forall n),~z_{n,l}\in\{0,1\}~ (\forall n,l).
\end{aligned}
\end{equation}
Problem~\eqref{mm_phase_milp} jointly updates all discrete phase coefficients and can be solved by a standard branch-and-bound MILP solver~\cite{bbmilp}. The binary representation of $\boldsymbol\phi^{(m)}$ is always feasible for~\eqref{mm_phase_milp}. Indeed, substituting $\boldsymbol\phi=\boldsymbol\phi^{(m)}$ into the left-hand side of~\eqref{milp_qos_constraint} gives
$2h_{t,k}(\boldsymbol\phi^{(m)})$, which is no smaller than
$\xi_{t,k}+h_{t,k}(\boldsymbol\phi^{(m)})$
because $\boldsymbol\phi^{(m)}$ is QoS-feasible. Thus, the MILP remains feasible at every iteration.

Let $\{z_{n,l}^{(m),\star}\}$ denote an optimal solution of~\eqref{mm_phase_milp}. The discrete phase angles are updated as
\begin{equation}
\label{mm_update}
\phi_n^{(m+1)}
\leftarrow
\sum_{l=1}^{L}u_lz_{n,l}^{(m),\star},
~
\theta_n^{(m+1)}
\leftarrow
\arg\big(\phi_n^{(m+1)}\big)~ (\forall n).
\end{equation}
with $\boldsymbol\phi^{(m+1)} = [e^{j\theta_1^{(m+1)}}\cdots e^{j\theta_N^{(m+1)}}]^{\rm T}$ and corresponding $\boldsymbol\theta^{(m+1)}$. Because~\eqref{milp_qos_constraint} is an inner approximation of the exact QoS constraint,~\eqref{h_phi_lb} implies
\begin{equation}
\label{phase_qos_preservation}
h_{t,k}(\boldsymbol\phi^{(m+1)})
\ge
\underline h_{t,k}
(\boldsymbol\phi^{(m+1)}|\boldsymbol\phi^{(m)})
\ge\xi_{t,k}~ (\forall k\in\mathcal C_t).
\end{equation}
Furthermore, since the current binary assignment is feasible for~\eqref{mm_phase_milp}, its solution cannot decrease the MM surrogate. Hence,
\begin{equation}
\label{phase_monotonic}
f(\boldsymbol\phi^{(m+1)})
\ge
\widetilde f
(\boldsymbol\phi^{(m+1)}|\boldsymbol\phi^{(m)})\ge
\widetilde f
(\boldsymbol\phi^{(m)}|\boldsymbol\phi^{(m)})
=f(\boldsymbol\phi^{(m)}).
\end{equation}
Therefore, every MM iteration preserves both the discrete hardware constraints and the original QoS constraints while generating a nondecreasing sum-rate sequence. Since $\mathcal F_b^N$ is finite, the procedure terminates after finite strict improvements, and the resulting $\{\boldsymbol\theta^{(m)}\}$ is hence a stationary point. 

\subsubsection{Update of $\boldsymbol\chi$}
For fixed $(\boldsymbol\theta,\boldsymbol\alpha,\boldsymbol\rho_t)$, $\boldsymbol\chi$ introduces a combinatorial search over $\binom{N}{N_o}$ possible element subsets. To efficiently solve this problem, we employ the CEM~\cite{CEM, FRISonoff, FARIS}, which is a stochastic optimization technique suitable for discrete optimization. Specifically, a probability distribution over the selection variables is iteratively updated based on elite samples that yield higher objective values. In each iteration, a set of candidate selection patterns is generated according to the current distribution, and the best-performing samples are used to update the distribution parameters.

Given $c$, let $\mathbf p^{(c)}=[p_1^{(c)} \cdots p_N^{(c)}]^{\mathrm T}$ denote the probability vector at CEM iteration $c$, where $p_n^{(c)}$ represents the probability that the $n$th element is selected. A set of FARIS element selection vectors $\{\boldsymbol\chi^{(m)}\}_{m=1}^{M}$ is generated according to $\mathbf p^{(c)}$, while enforcing $\sum_{n=1}^{N}\chi^{(n)}=N_o$. For each $\boldsymbol\chi^{(m)}$, the corresponding $\{\boldsymbol\alpha^{(m)}\}$ is optimized by Section~\ref{sasa}, the corresponding objective value in~\eqref{p1} is evaluated\footnote{Since by optimization in Section~\ref{sasa}, the QoS is satisfied for all samples.}, and the top-performing $M_e$ samples are selected as the elite set $\mathcal E^{(c)}$. The probability vector $\mathbf p^{(c)}$ is then updated according to the empirical selection frequency of the elite samples as $p_n^{(c+1)}=\frac{1}{M_e}\sum_{m\in\mathcal E^{(c)}} \chi_n^{(m)}~(\forall n)$. To improve numerical stability and avoid premature convergence, a smoothing step is adopted: $\mathbf p^{(c+1)}
\leftarrow
(1-\alpha_{\mathrm{CE}})\mathbf p^{(c)}+\alpha_{\mathrm{CE}}\mathbf p^{(c+1)}$, where $\alpha_{\mathrm{CE}}\in(0,1]$ denotes the CEM smoothing parameter. After $\{\mathbf p^{(c)}\}$ converges, the corresponding selection pattern $ \boldsymbol\chi^\star$ is obtained by selecting the $N_o$ FARIS elements with the largest selection probabilities: $\chi^{*(n)} =
\begin{cases}
1 & (n \in \mathcal S^\star) \\
0 & (\text{otherwise})
\end{cases}$, where $\mathcal S^\star$ denotes the index set corresponding to the $N_o$ largest entries of the converged probability vector. 
As the probability distribution becomes increasingly concentrated around high-performing selection patterns~\cite{CEM}, the CEM iteratively improves the objective value and converges to a near-optimal FARIS configuration.

\subsection{Overall AO Framework}
\label{subsec:overall_ao}
The overall procedure for the $t$th cluster is summarized in Algorithm~\ref{alg:AO_FARIS_NOMA} with AO iteration index $\ell$. For a given \(\mathcal C_t\), the AO procedure starts from a feasible tuple \((\boldsymbol\rho_t^{(0)},\boldsymbol\alpha^{(0)},\boldsymbol\theta^{(0)},\boldsymbol\chi^{(0)})\). Herein, \(\boldsymbol\chi^{(0)}\) and \(\boldsymbol\theta^{(0)}\) are randomly generated subject to $\sum_{n=1}^{N}\chi_n^{(0)}=N_o$ and $(\rho_{t,k}^{(0)}\ge0, \sum_{k\in\mathcal C_t}\rho_{t,k}^{(0)}=1)$, respectively, and each entry of \(\boldsymbol\theta^{(0)}\) is independently selected from \(\mathcal Q_b\). Finally, by letting $\widehat{\boldsymbol\alpha}^{(0)}
=
\alpha_{\max}\boldsymbol\chi^{(0)}$, $\boldsymbol\alpha^{(0)}$ is obtained by scaling \(\widehat{\boldsymbol\alpha}^{(0)}\) as $\boldsymbol\alpha^{(0)}
=
\min\left\{
1,
\sqrt{
\frac{P_{\max}}
{
(\widehat{\boldsymbol\alpha}^{(0)})^{\mathrm T}
\mathbf C_t(\boldsymbol\chi^{(0)})
\widehat{\boldsymbol\alpha}^{(0)}
}
}
\right\}\widehat{\boldsymbol\alpha}^{(0)}$. Since each inner block is optimized while keeping the others fixed, the objective value of~\eqref{p1} is nondecreasing over $\ell$. 

\begin{algorithm}[t]
\caption{Proposed AO Framework Per Cluster}
\label{alg:AO_FARIS_NOMA}
\begin{algorithmic}[1]
\Require System parameters, 
AO tolerance $\varepsilon>0$.
\State \textbf{Initialization:} Feasible $(\boldsymbol\rho_t^{(0)},\boldsymbol\alpha^{(0)},\boldsymbol\theta^{(0)},\boldsymbol\chi^{(0)})$, $\ell\leftarrow 0$, $\mathcal R_t^{(0)}=\mathcal R_t(\boldsymbol\rho_t^{(0)},\boldsymbol\alpha^{(0)},\boldsymbol\theta^{(0)},\boldsymbol\chi^{(0)})$.

\While{$|\mathcal R_t^{(\ell+1)}-\mathcal R_t^{(\ell)}|\ge \varepsilon$}
    \Statex \textbf{1) Update $\boldsymbol\rho_t$ for fixed $(\boldsymbol\alpha^{(\ell)},\boldsymbol\theta^{(\ell)},\boldsymbol\chi^{(\ell)})$: $\boldsymbol\rho_t^{(\ell+1)}$}

    \Statex \textbf{2) Update $\boldsymbol\alpha$ for fixed $(\boldsymbol\rho_t^{(\ell+1)},\boldsymbol\theta^{(\ell)},\boldsymbol\chi^{(\ell)})$: $\boldsymbol\alpha^{(\ell+1)}$}

    \Statex \textbf{3) Update $\boldsymbol\theta$ for fixed $(\boldsymbol\rho_t^{(\ell+1)},\boldsymbol\alpha^{(\ell+1)},\boldsymbol\chi^{(\ell)})$: $\boldsymbol\theta^{(\ell+1)}$}
    
    \Statex \textbf{4) Update $\boldsymbol\chi$ for fixed $(\boldsymbol\rho_t^{(\ell+1)},\boldsymbol\alpha^{(\ell+1)},\boldsymbol\theta^{(\ell+1)})$: $\boldsymbol\chi^{(\ell+1)}$}

    \State $    \mathcal R_t^{(\ell+1)}
    =
    \mathcal R_t(\boldsymbol\rho_t^{(\ell+1)},\boldsymbol\alpha^{(\ell+1)},\boldsymbol\theta^{(\ell+1)},\boldsymbol\chi^{(\ell+1)})$.
    \State $\ell\leftarrow \ell+1$.
\EndWhile

\State \Return $(\boldsymbol\rho_t^\star,\boldsymbol\alpha^\star,\boldsymbol\theta^\star,\boldsymbol\chi^\star)\leftarrow(\boldsymbol\rho_t^{(\ell)},\boldsymbol\alpha^{(\ell)},\boldsymbol\theta^{(\ell)},\boldsymbol\chi^{(\ell)})$.
\end{algorithmic}
\end{algorithm}

\subsection{Computational Complexity Analysis}
\label{subsec:complexity}
We analyze the computational complexity of Algorithm~\ref{alg:AO_FARIS_NOMA}. Let $\mathcal C_{\rm QCQP}(n,m;\varepsilon)$ and $\mathcal C_{\rm LP}(n,m;\varepsilon)$ denote the complexity of solving a convex QCQP and LP, respectively, with $n$ real decision variables and $m$ constraints to accuracy $\varepsilon$. For a generic dense primal-dual interior-point implementation, conservative estimates are~\cite{boyd}:
\begin{equation}
\label{solver_complexities}
\begin{aligned}
\mathcal C_{\rm QCQP}(n,m;\varepsilon)
&=
\mathcal O\left(
\sqrt{m}\log(1/\varepsilon)(n+m)^3
\right),\\
\mathcal C_{\rm LP}(n,m;\varepsilon)
&=
\mathcal O\left(
\sqrt{m}\log(1/\varepsilon)(n+m)^3
\right),
\end{aligned}
\end{equation}

\subsubsection{Update of $\boldsymbol\rho_t$}
The GP in~\eqref{gp_rho} contains $2K_t$ positive optimization variables, i.e., $\{\rho_{t,k},\tau_{t,k}\}_{k\in\mathcal C_t}$, and is solved iteratively using the AGMA-based monomial approximation. Let $I_{\rho}$ denote the number of inner GP iterations. As each approximated GP can be transformed into a convex problem in the log-domain and solved by an interior-point method, the complexity of the $\boldsymbol\rho_t$-update is expressed as $\mathcal C_{\rho}
=
\mathcal O\left(I_{\rho}K_t^3\right)$.
\subsubsection{Update of $\boldsymbol\alpha$}
At each inner FP/SCA iteration, $\{\lambda_{t,k},y_{t,k}\}_{k\in\mathcal C_t}$
are updated in closed-form according to~\eqref{lambda_update} and~\eqref{y_update}. By exploiting the rank-one-plus-diagonal structures of
$\mathbf S_{t,k}$, $\mathbf N_{t,k}$, and $\mathbf T_{t,k}$, the required quadratic forms can be evaluated in
$\mathcal O(K_tN)$ operations. Forming the dense aggregate matrix
$\mathbf M_t$ and the $K_t$ QoS quadratic constraints requires at most
$\mathcal O(K_tN^2)$ operations. 
The QCQP in~\eqref{pa3} has $N$ real decision variables, $2N$ box constraints, one FARIS power constraint, and $K_t$ QoS constraints. Hence, with $m_{\alpha}\triangleq2N+K_t+1$, the cost of one QCQP solve is
$\mathcal C_{\rm QCQP}(N,m_{\alpha};\varepsilon_{\alpha})$.
Let $I_{\alpha}$ denote the number of inner FP/SCA iterations. The overall cost of $\boldsymbol\alpha$-update is therefore $\mathcal C_{\alpha}
=
\mathcal O\left(
I_{\alpha}
\left[
K_tN^2
+\mathcal C_{\rm QCQP}
(N,m_{\alpha};\varepsilon_{\alpha})
\right]
\right)$.

\subsubsection{Update of $\boldsymbol\theta$}
Let $L\triangleq2^b$ denote the number of available phase coefficients. At each discrete MM iteration, evaluating~\eqref{grad_phi} costs
$\mathcal O(K_tN)$ since
$\mathbf Q_{t,k}=\mathbf d_{t,k}\mathbf d_{t,k}^*$
is rank-one. Constructing 
$\{c_{n,l}^{(m)}\}$ and
$\{d_{t,k,n,l}^{(m)}\}$ requires
$\mathcal O(K_tNL)$ operations. The MILP in~\eqref{mm_phase_milp} contains $n_{\theta}=NL=N2^b$ binary variables, $N$ one-hot equality constraints, and $K_t$ QoS inequalities. Let
$m_{\theta}\triangleq N+K_t$
and let $B_{\theta}^{(m)}$ denote the number of branch-and-bound nodes explored when solving the MILP at the $m$th MM iteration. Each node requires solving an LP relaxation whose cost is represented by
$\mathcal C_{\rm LP}(n_{\theta},m_{\theta};\varepsilon_{\theta})$.
Accordingly, letting
$\bar B_{\theta}\triangleq
I_{\theta}^{-1}\sum_{m=1}^{I_{\theta}}B_{\theta}^{(m)}$
denote the average number of explored nodes and $I_{\theta}$ the number of discrete MM iterations, the phase-update complexity is $\mathcal C_{\theta}
=
\mathcal O\left(
I_{\theta}
\left[
K_tN2^b
+\bar B_{\theta}
\mathcal C_{\rm LP}
(N2^b,N+K_t;\varepsilon_{\theta})
\right]
\right)$.

\subsubsection{Update of $\boldsymbol\chi$}
At each CEM iteration, $M$ feasible element-selection candidates are retained. For every sampled $\boldsymbol\chi^{(m)}$, the corresponding $\boldsymbol\alpha^{(m)}$ is obtained using the QoS-preserving amplification-gain procedure in Section~\ref{sasa}. Consequently, the dominant cost of evaluating one candidate is the optimization of $\{\boldsymbol\alpha^{(m)}\}$, whose complexity is $M\mathcal C_{\alpha}$, followed by an additional $\mathcal O(MK_tN)$ evaluation of the effective channels, SINRs, and cluster sum-rate. Generating the fixed-cardinality samples and updating the probability vector require
$\mathcal O(MN)$ operations, whereas sorting the retained objective values requires
$\mathcal O(M\log M)$. Thus, with $I_{\rm CEM}$ CEM iterations, the element-selection update has complexity $\mathcal C_{\chi}
=
\mathcal O\Big(
I_{\rm CEM}
\big[
M\mathcal C_{\alpha}
+MN+M\log M
\big]
\Big)$.

\subsubsection{Total Computational Complexity}
Hence with $I_{\rm AO}$ iterations, the computational complexity of Algorithm~\ref{alg:AO_FARIS_NOMA} is
\begin{equation}
\label{eq:complexity_total}
\begin{aligned}
\mathcal O\Bigg(
I_{\rm AO}
\Bigg[&
I_{\rho}K_t^3
+\mathcal C_{\alpha}+I_{\theta}
\left\{
K_tN2^b
+\bar B_{\theta}
\mathcal C_{\rm LP}
(N2^b,N+K_t;\varepsilon_{\theta})
\right\}\\
&+I_{\rm CEM}
\left\{
M\mathcal C_{\alpha}+MN+M\log M
\right\}
\Bigg]
\Bigg).
\end{aligned}
\end{equation}
For all $C$ clusters, the total cost is obtained by summing~\eqref{eq:complexity_total} over $t=1,\cdots,C$.

\section{Simulation Results}
\begin{figure}[t]
	\begin{center}
		\includegraphics[width=0.5\columnwidth,keepaspectratio]%
		{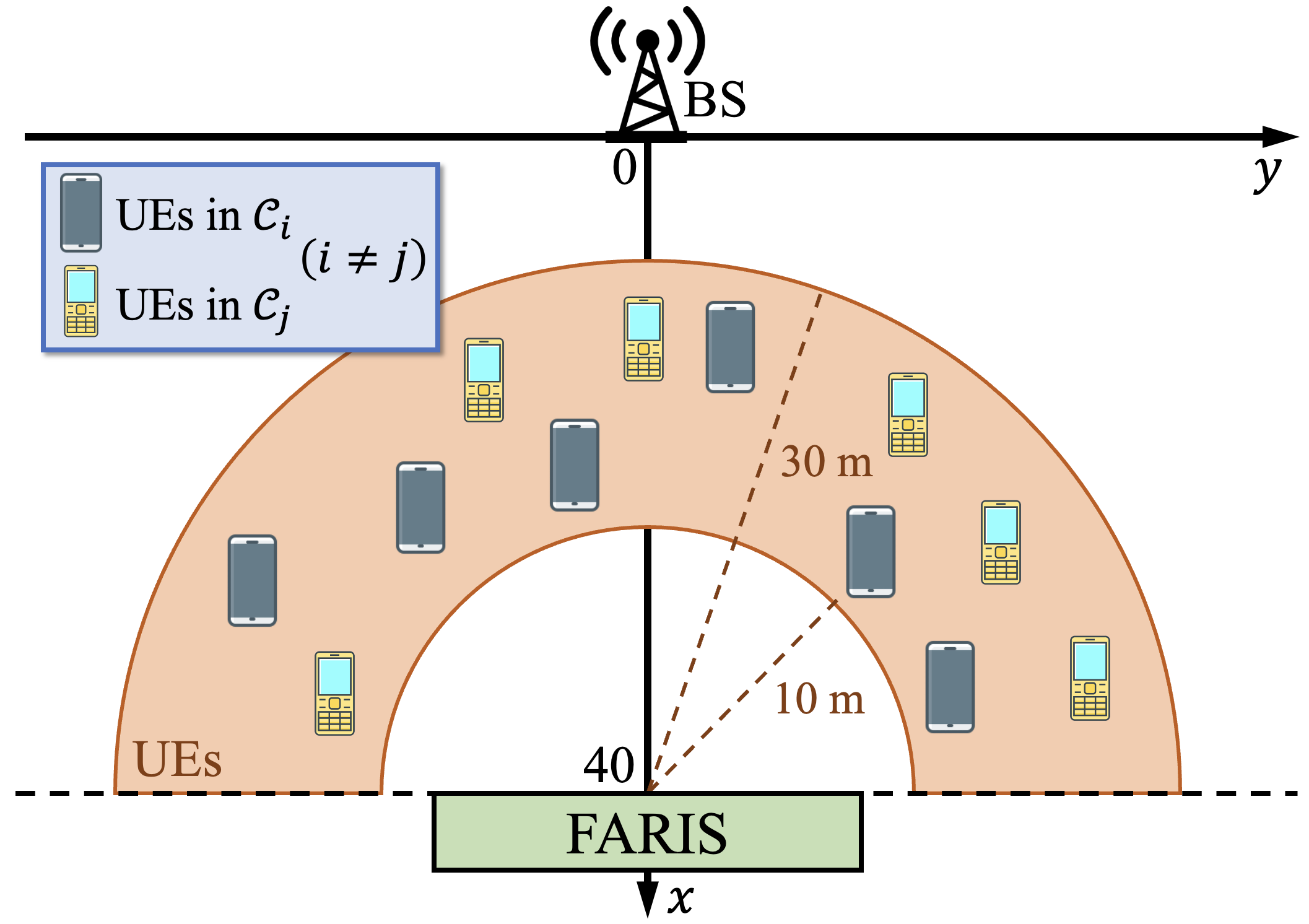}
		\caption{Simulation setup of the FARIS-aided downlink NOMA system.}
		\label{fig_setsim}
	\end{center}
\end{figure}

We evaluate the proposed FARIS-aided downlink NOMA design for both per-cluster and network-level settings with 1,000 Monte Carlo simulations. The BS and FARIS are fixed at $[0, 0, 10]$ and $[40, 0, 5]$~m, respectively, while the UEs are randomly distributed in the range of 10-30~m from the utilized-side of the FARIS to create sufficient near-far channel disparity for NOMA clustering~\cite{FRISnoma}. The illustration of the simulation environment is depicted in Fig.~\ref{fig_setsim}, and the detailed parameters are given in Table~\ref{tab:sim_param}. For the network-level evaluation, the UEs are first grouped into $C$ clusters using the distance-based interleaved clustering in Section~\ref{noc}, and the per-cluster optimization is then carried out independently over orthogonal time slots. The aggregated network sum-rate is then obtained by summing the optimized per-cluster rates over the $C$ orthogonal time slots.
\begin{table}[t]
\centering
\caption{Simulation Parameters (Unless Otherwise Stated)}
\label{tab:sim_param}
\begin{tabular}{c c}
\toprule
\textbf{Parameter} & \textbf{Value} \\ 
\midrule
PL parameters $(\tau, \zeta)$ &  (-46 [dB], 2.8)\\ 
Transmit and FARIS power budget $(P_t, P_{\max,t})$ & (20, 25)~[dBm] \\ 
Noise power $(\sigma_\eta^2, \sigma_F^2)$ & -90~[dBm] (same) \\ 
Hardware-related power $(P_c, P_{\mathrm{DC}})$ & (-10, -5)~[dBm]~\cite{aris5} \\ 
Maximum amplification gain $\alpha_{\max}$ &  40~[dB]~\cite{aris5} \\
Phase resolution $b$ & 3~[bits] \\ 
Minimum rate $R_{\mathrm{th}}$ & 0.5~[bps/Hz] \\
Normalized FARIS size $W_x$ & 2 \\ 
Number of FARIS candidates $N$ & 64 ($8\times8$) \\ 
Number of selected FARIS elements $N_o$ & 16 \\ 
Number of users $U$ & 48 \\ 
Number of clusters $C$ & 12 \\ 
Parameters in CEM $(N_{\mathrm{mc}}, \rho, \alpha_{\rm CE})$ & $(5N, 0.1, 0.7)$~\cite{CEM}\\
Every inner/outer AO tolerance & $10^{-3}$\\
\bottomrule
\end{tabular}
\end{table}

As benchmarks, we consider three representative baselines:
\begin{itemize}
\item \textbf{FARIS-OMA}: We employ the same FARIS architecture but it serves users via OMA. Specifically, $(\boldsymbol\alpha,\boldsymbol\theta,\boldsymbol\chi)$ are optimized using the proposed AO framework, where the transmit power is equally divided among users.
\item \textbf{FRIS-NOMA}: We retain the fluid port-selection capability but remove active amplification~\cite{FRISnoma}. 
\item \textbf{ARIS-NOMA}: We adopt a conventional ARIS structure with the same number of $N_o$ elements. In this case, $(\boldsymbol\rho_t, \boldsymbol\alpha,\boldsymbol\theta)$ are optimized under the same AO framework and power constraint.
\end{itemize}

\subsection{Reliability of Proposed Framework}
\label{rpf}
\begin{figure}[t]
    \centering
    \subfloat[]{%
        \includegraphics[width=0.25\textwidth]{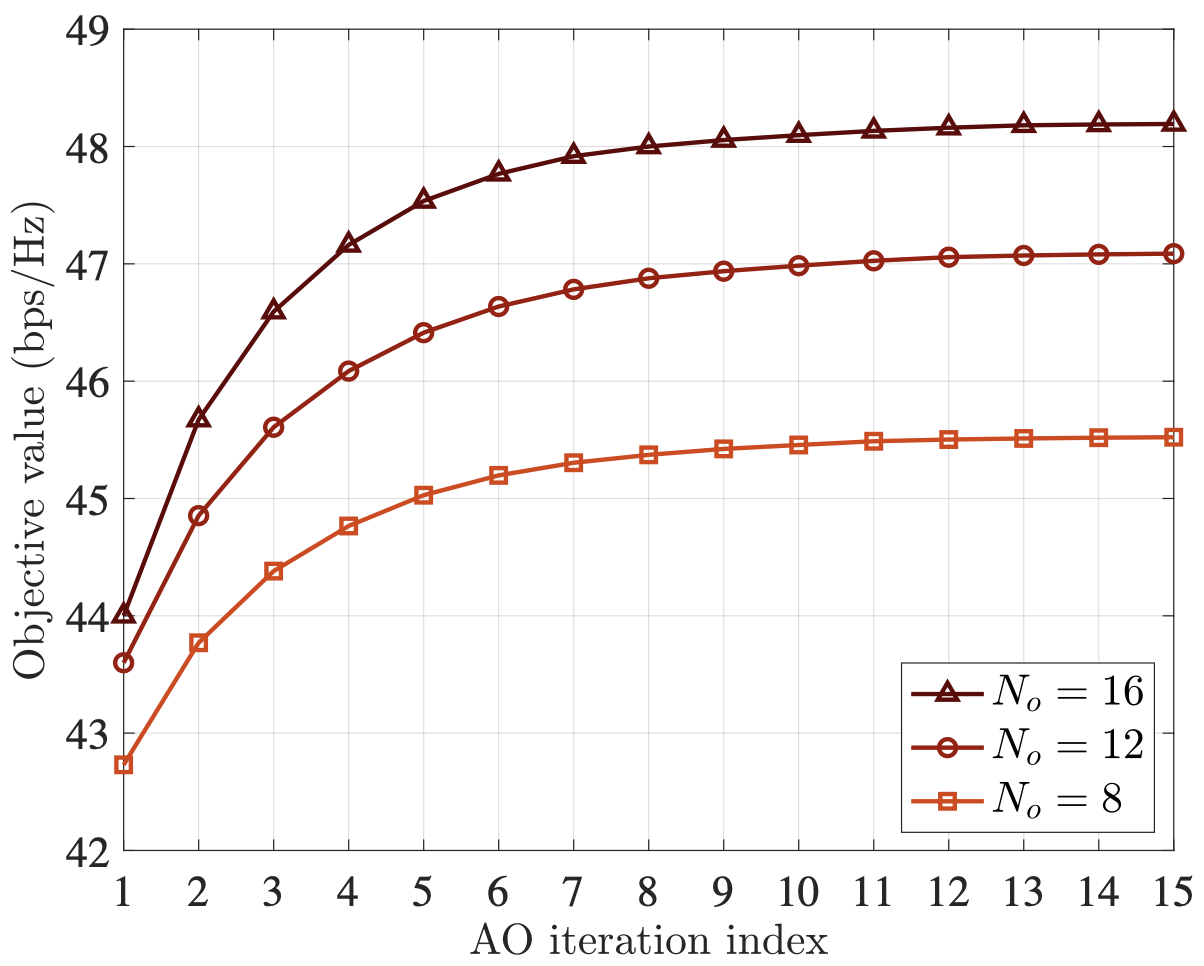}
        \label{fig_conv}%
    }
    \vfil
    \subfloat[]{%
        \includegraphics[width=0.48\textwidth]{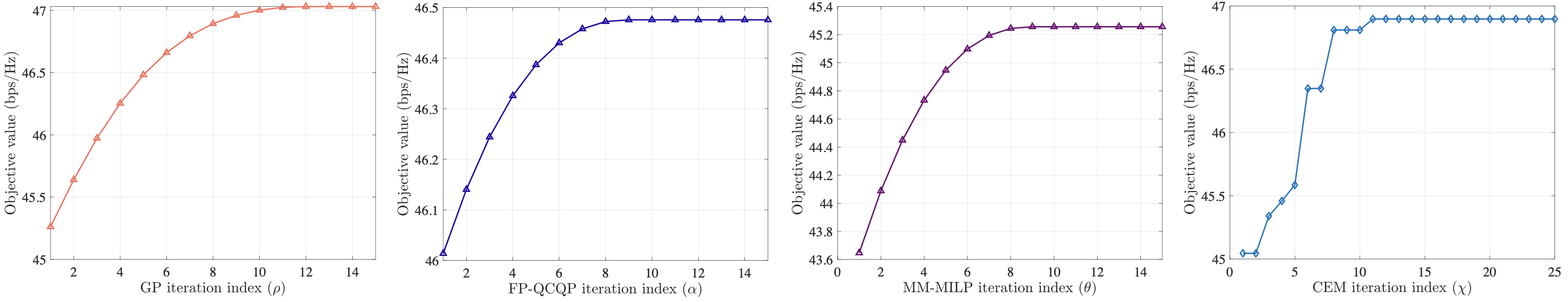}
        \label{fig_convin}%
    }
    \caption{Convergence of the proposed AO framework with respect to (a) AO outer iteration index for different $N_o$ and (b) subproblem inner indices.}
    \label{fig_2}
\end{figure}
To validate the convergence of the proposed AO framework, Figs.~\ref{fig_conv} and~\ref{fig_convin} plot the objective value of~\eqref{p0} versus the AO outer iteration index for different $N_o$ and the inner iteration indices of the subproblems, respectively. 
As observed, the objective value increases monotonically and converges rapidly in both cases, confirming that each block update effectively improves the system performance. In particular, most of the gain is achieved within the first few iterations, after which the objective quickly stabilizes. This behavior is consistent with the proposed design, since each subproblem is handled to yield a nondecreasing objective sequence at every step. The smooth and fast convergence further demonstrates the stability and computational efficiency of the proposed framework, which will be compared with the BFS method in the followings.


\begin{figure}[t]
	\begin{center}
		\includegraphics[width=0.5\columnwidth,keepaspectratio]%
		{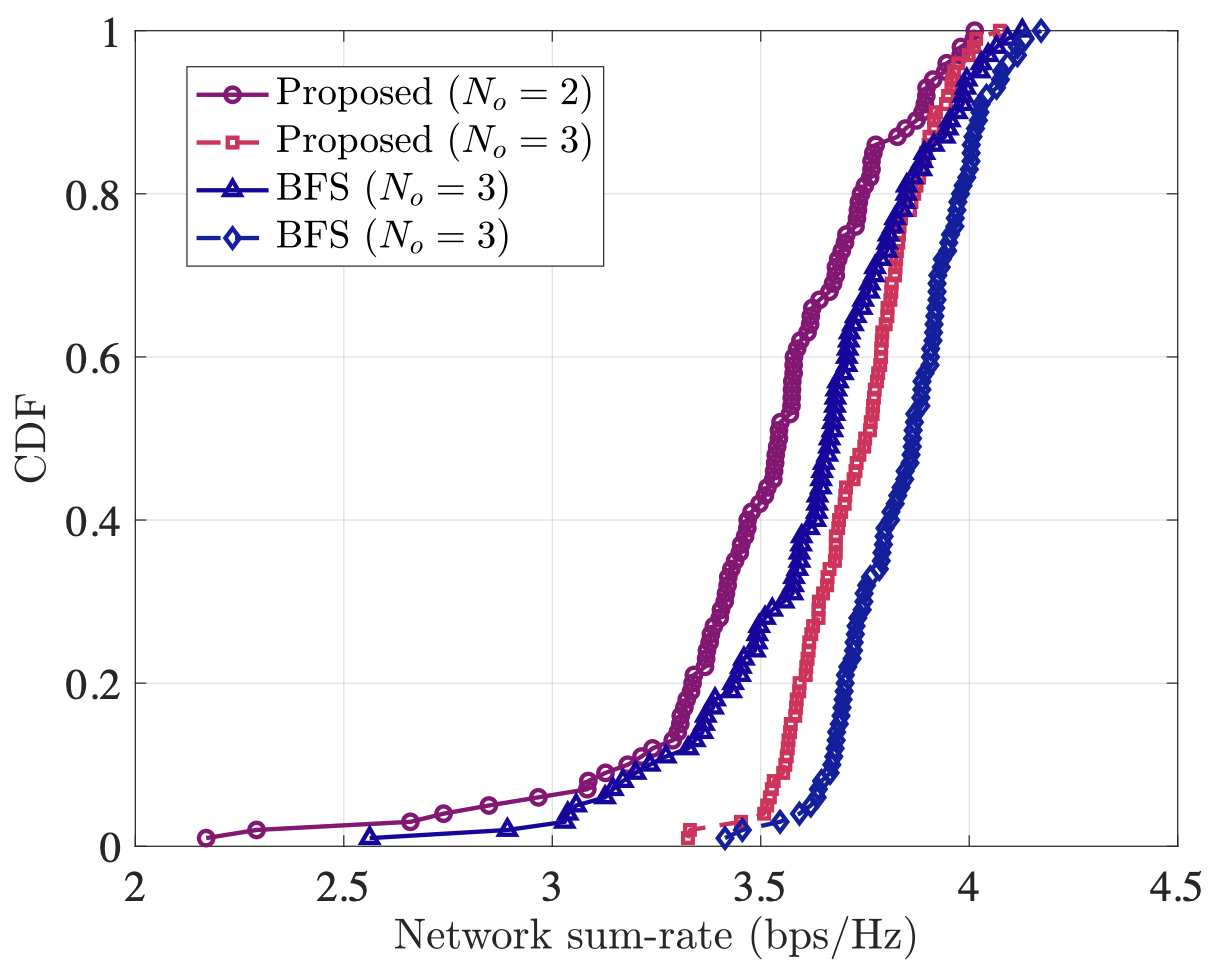}
\caption{Empirical CDF of $\mathcal R$ for the proposed algorithm and the brute-force search (BFS) benchmark under a small-scale setting with $N=4$, $N_o\in\{2,3\}$, $b=2$, and $(U, C)=(6, 3)$, whereas continuous \(\boldsymbol\rho_t\) and $\boldsymbol\alpha$ and are herein discretized by order of $\frac{1}{10}$ and 2~[dB], respectively.}
		\label{fig_cdf}
	\end{center}
\end{figure}
Fig.~\ref{fig_cdf} evaluates the optimality of the proposed algorithm by comparing its performance with the BFS method via the empirical cumulative distribution function (CDF) of $\mathcal R$. Herein, the BFS provides the almost-global optimal solution of~\eqref{p0} by exhaustively searching all feasible configurations, which is feasible in a given relatively small-scale setting. As shown in the figure, the CDF curves of the proposed algorithm closely approach those of the BFS for both $N_o=4$ and $6$; in particular, the average performance achieves approximately $94.62\%$ and $95.17\%$ of the BFS benchmark 
for $N_o=4$ and $6$, respectively, demonstrating that the proposed method achieves near-optimal performance. Although a slight performance gap is observed due to the slight sub-optimality of the proposed approach, the gap remains marginal across the entire distribution. Moreover, the relative ordering of the curves confirms that increasing $N_o$ improves the achievable sum-rate for both. These results verify that the proposed AO framework provides an effective approximation to the optimal solution while significantly reducing the computational burden compared to exhaustive search.

\subsection{Sum-Rate Comparisons under Several Effects}
\label{beco}
\begin{figure}[t]
	\begin{center}
		\includegraphics[width=0.5\columnwidth,keepaspectratio]%
		{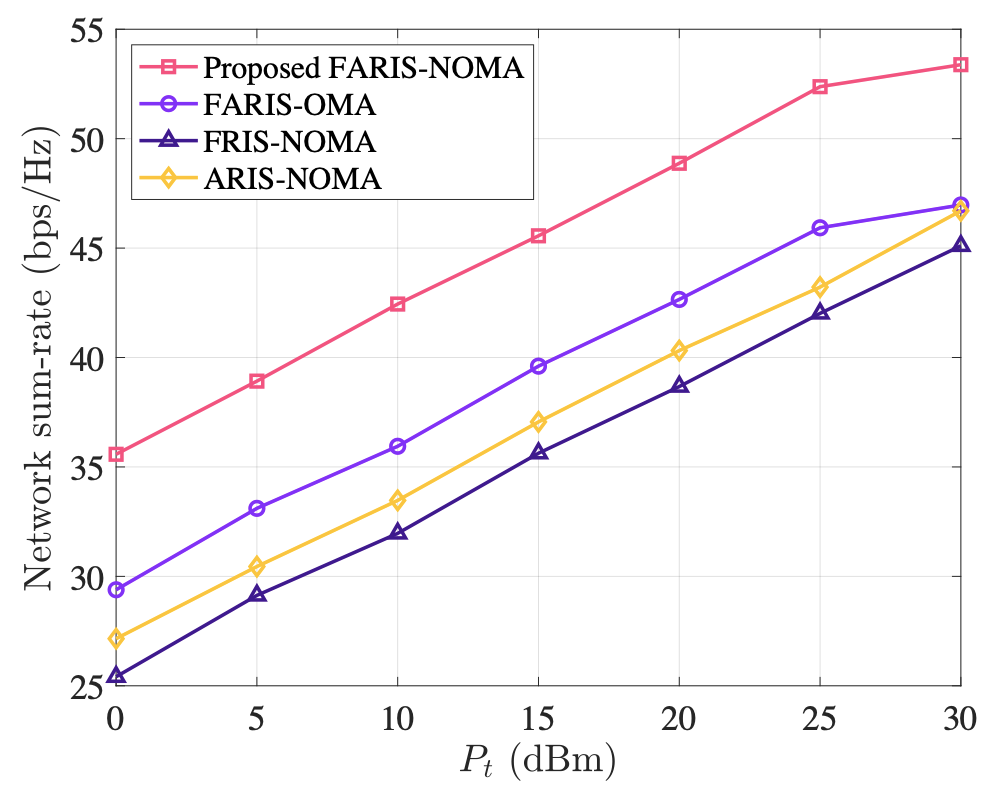}
\caption{Network sum-rate versus $P_t$ for the proposed FARIS-NOMA and benchmarks.}
		\label{fig_pt}
	\end{center}
\end{figure}
Fig.~\ref{fig_pt} illustrates the network sum-rate versus $P_t$ for different schemes. From the figure, the proposed FARIS-NOMA consistently achieves the highest sum-rate across the entire power range; $16.5\sim30.3\%$ increase compared to baselines at $P_t=20$~dBm. This performance gain stems from the proposed joint optimization of fluid port selection, active amplification, and NOMA transmission, which effectively enhances the cascaded channel gain and improves spectral efficiency. In contrast, FARIS-OMA suffers from the lack of power-domain multiplexing despite using the same hardware architecture, resulting in a noticeable performance gap. Moreover, FRIS-NOMA shows inferior performance due to the absence of active amplification, while ARIS-NOMA is limited by its static element configuration without fluid reconfiguration capability. These results indicate that the proposed design more effectively converts the increased $P_t$ into useful signal enhancement through joint optimization, thereby achieving higher scalability and spectral efficiency.

\begin{figure}[t]
	\begin{center}
		\includegraphics[width=0.5\columnwidth,keepaspectratio]%
		{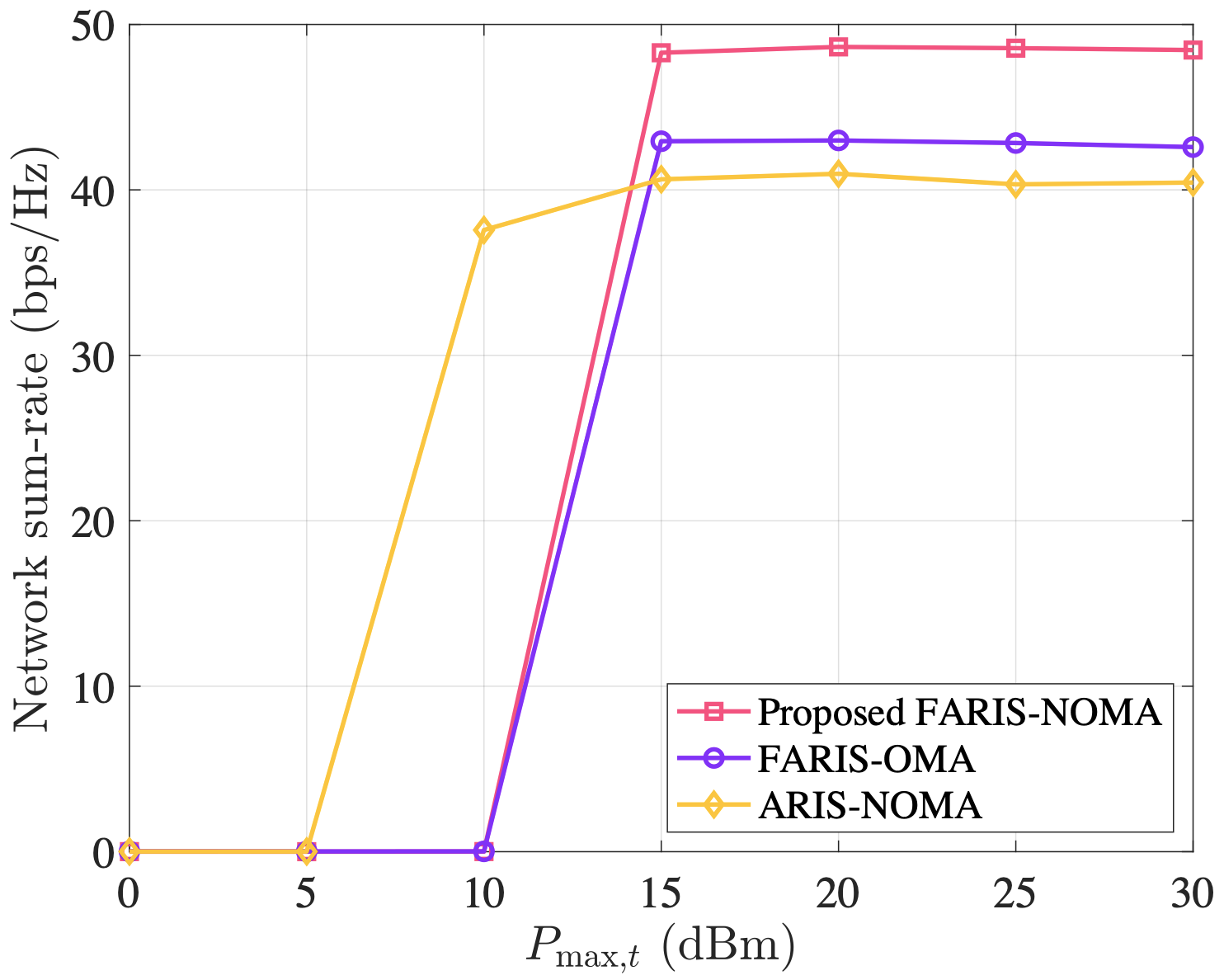}
\caption{Network sum-rate versus $P_{\max,t}$ for the proposed FARIS-NOMA and benchmarks.}
		\label{fig_pmax}
	\end{center}
\end{figure}
Fig.~\ref{fig_pmax} illustrates the network sum-rate versus $P_{\max,t}$ for amplification-related systems. Herein, all schemes achieve zero or negligible rates in the low-power regime, where the reflected-power constraint severely limits the effective signal strength~\cite{aris4}. As $P_{\max,t}$ increases around $P_{\max,t}\approx 10\sim 15$~dBm, the sum-rate rapidly improves, indicating a transition from a power-limited regime to an operation region where sufficient amplification is available. This transition occurs at a slightly larger $P_{\max,t}$ than that of ARIS (around $5$~dBm), which stems from the adopted FARIS hardware power model in~\eqref{eq:Pfaris_total} and corresponding~\eqref{eq:Pris_constraint}. In particular, since FARIS must reserve a portion of the total power budget for the logical control power ($P_c$) of all $N$ candidate ports and the DC bias power ($P_{\mathrm{DC}}$) of the $N_o$ selected ports, the effective reflected-power budget in~\eqref{eq:Pris_constraint} becomes smaller for a given $P_{\max,t}$. As a result, a larger total power budget is required before sufficient radiated reflection power becomes available, thereby shifting the onset of the rapid sum-rate increase to a higher $P_{\max,t}$ region. Nevertheless, the proposed FARIS-NOMA exhibits the fastest growth and achieves the highest performance in this region of $P_{\max,t}$; $16.5\sim20.7\%$ increase compared to benchmarks, thanks to the proposed AO framework. In particular, ARIS-NOMA suffers from its fixed element configuration, leading to the saturation with lower rate, while FARIS-OMA cannot exploit NOMA gains despite hardware flexibility. This demonstrates the effectiveness of the proposed framework in translating power budget into meaningful rate gains.

\begin{figure}[t]
	\begin{center}
		\includegraphics[width=0.5\columnwidth,keepaspectratio]%
		{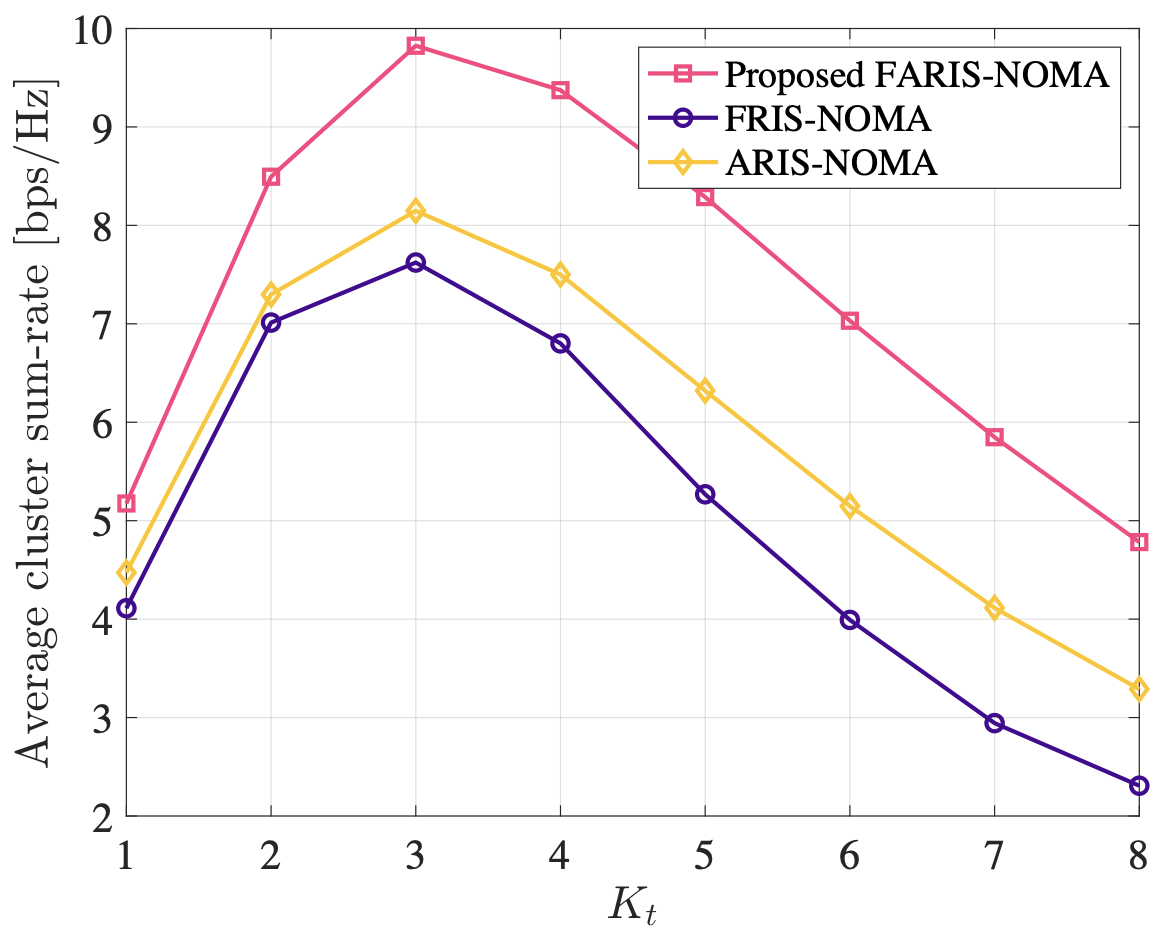}
\caption{Average cluster sum-rate versus $K_t$ for the proposed FARIS-NOMA and benchmarks.}
		\label{fig_kt}
	\end{center}
\end{figure}
Fig.~\ref{fig_kt} illustrates the average cluster sum-rate versus $K_t$ for NOMA-related systems, where $K_t$ is same for all $t$. It can be observed that the average cluster sum-rate initially increases as $K_t$ grows from a small value. However, beyond a moderate cluster size; herein $K_t\approx 3\sim4$, since a larger $K_t$ intensifies the intra-cluster interference and reduces the effective power available to each user, the sum-rate starts to decrease~\cite{FRISnoma, nomabehav}. Nevertheless, the proposed FARIS-NOMA consistently achieves the highest performance over the entire range of $K_t$, providing $36.6\sim76.1\%$ increase over the benchmark schemes at $K_t=6$. This demonstrates that the proposed joint design more effectively mitigates the performance degradation caused by large $K_t$, by better managing intra-cluster interference and power allocation as $K_t$ increases.

\begin{figure}[t]
	\begin{center}
		\includegraphics[width=0.5\columnwidth,keepaspectratio]%
		{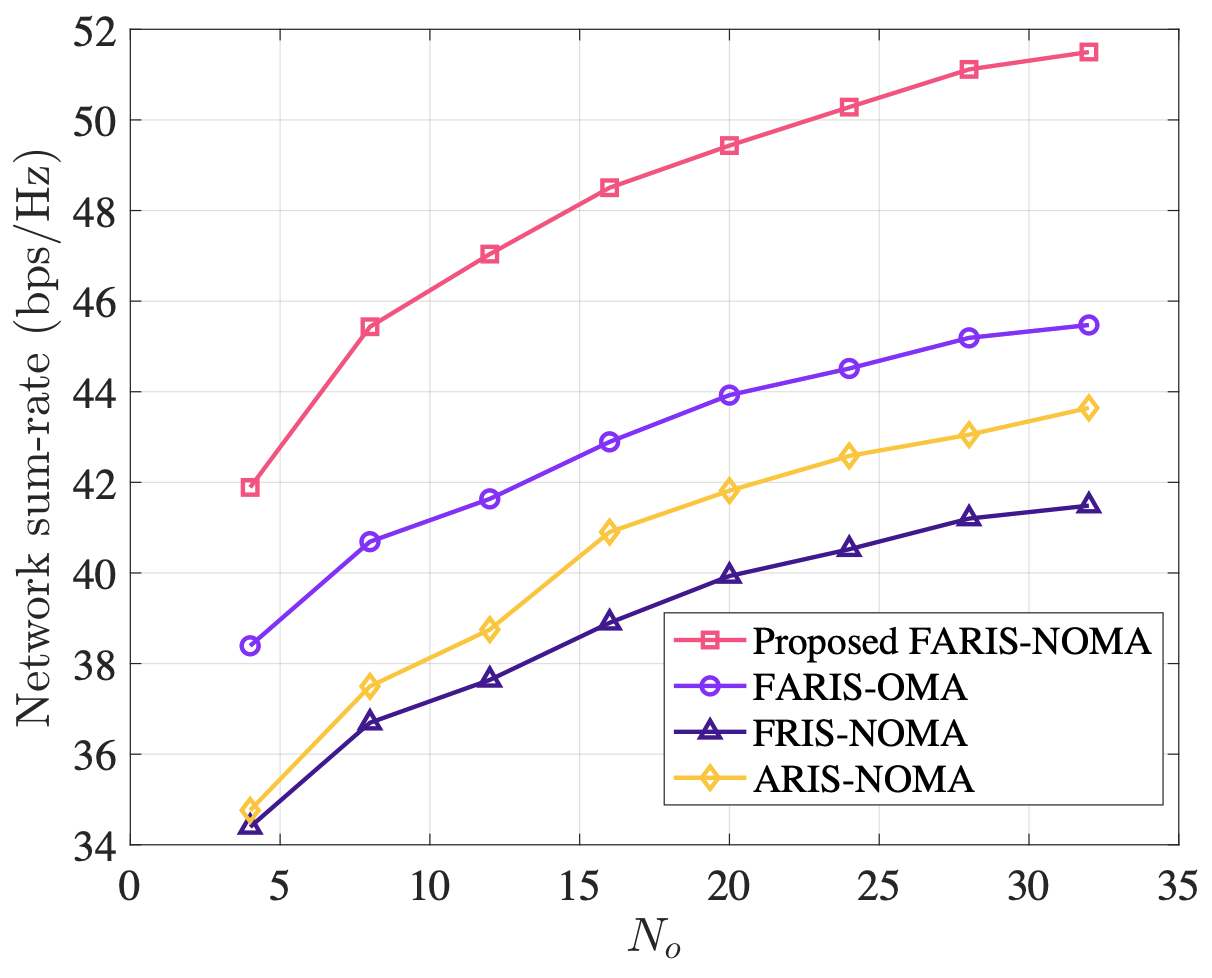}
\caption{Network sum-rate versus $N_o$ for the proposed FARIS-NOMA and benchmarks.}
		\label{fig_no}
	\end{center}
\end{figure}

Fig.~\ref{fig_no} illustrates the network sum-rate versus $N_o$. Increasing $N_o$ improves the sum-rate for all schemes, as activating more elements directly enhances the effective aperture and beamforming capability of the surface~\cite{FARIS, FRISlook}. The proposed FARIS-NOMA consistently outperforms the benchmark schemes by leveraging the proposed AO framework, whereas the benchmarks are constrained by either the absence of amplification or limited structural adaptability; $13.4\sim24.7\%$ increase compared to benchmarks when $N_o=24$. Thereafter, the performance improvement becomes marginal when $N_o$ approaches $N$, since the additional gain from activating more elements diminishes due to increased spatial correlation and the finite spatial DoF of the aperture.

\begin{figure}[t]
	\begin{center}
		\includegraphics[width=0.5\columnwidth,keepaspectratio]%
		{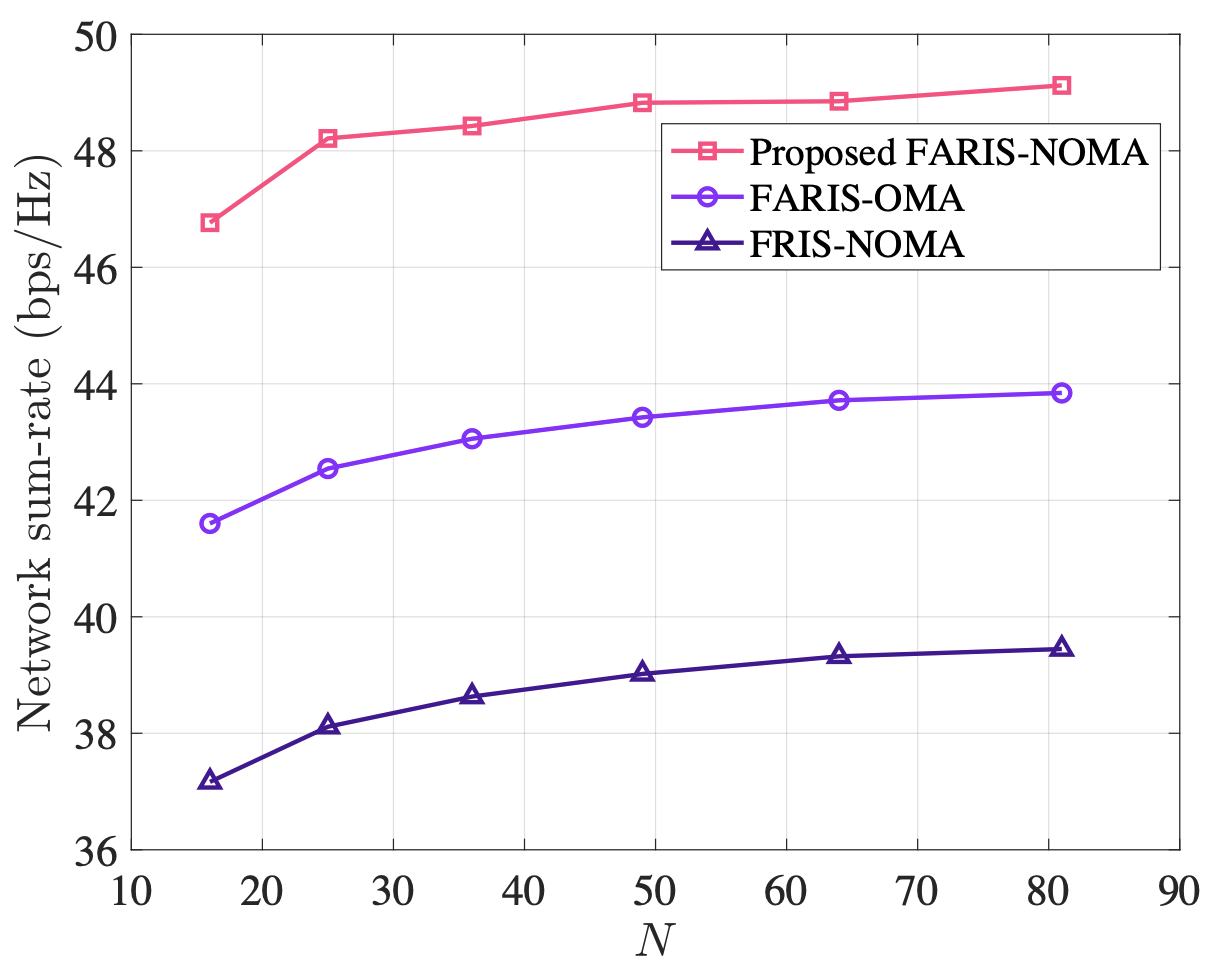}
\caption{Network sum-rate versus $N$ for the proposed FARIS-NOMA and benchmarks.}
		\label{fig_nn}
	\end{center}
\end{figure}

Fig.~\ref{fig_nn} presents the network sum-rate as a function of $N$ for fluidity-enabled frameworks. Herein, increasing $N$ enlarges the candidate pool, which provides greater flexibility in selecting favorable element positions and enables more efficient surface reconfiguration~\cite{FARIS, FRISsec}. As a result, all schemes benefit from the increased spatial diversity and improved configuration capability. The proposed FARIS-NOMA achieves the highest performance; $16.1\sim23.4\%$ increase compared to baselines when $N=81~(9\times9)$; by effectively exploiting this expanded design space through our proposed framework with NOMA transmission, thereby yielding superior utilization of the available spatial resources compared to the benchmarks.

\begin{figure}[t]
	\begin{center}
		\includegraphics[width=0.5\columnwidth,keepaspectratio]%
		{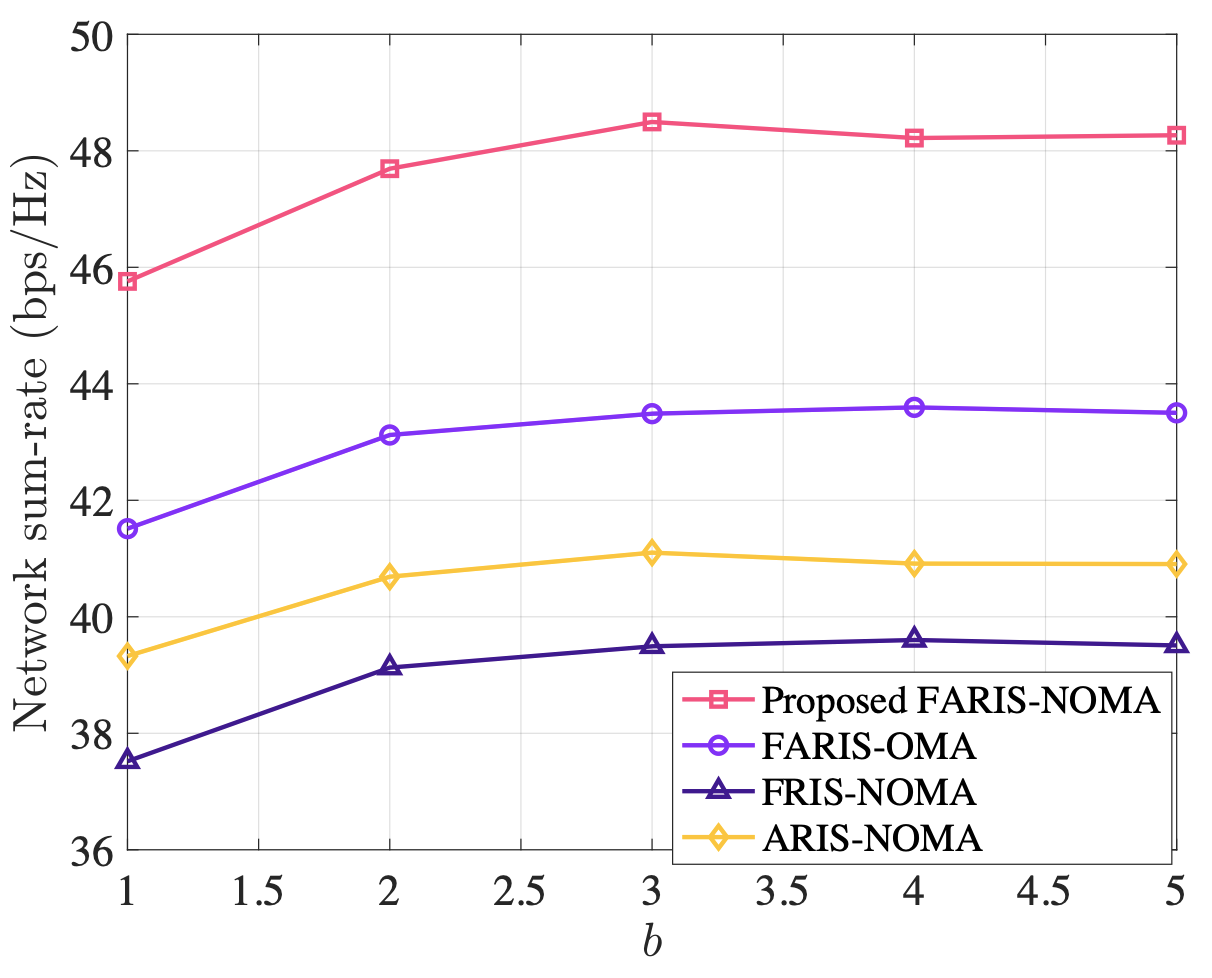}
\caption{Network sum-rate versus $b$ for the proposed FARIS-NOMA and benchmarks.}
		\label{fig_bb}
	\end{center}
\end{figure}

Fig.~\ref{fig_bb} illustrates the network sum-rate versus $b$. As $b$ increases, the sum-rate increases since finer phase quantization enables more accurate beamforming and reduces phase mismatch. However, the gain gradually saturates as $b$ becomes large, since the phases approach continuous values and further quantization refinement yields diminishing returns. Nevertheless, the proposed FARIS-NOMA consistently achieves the highest performance compared to the benchmarks.

\begin{figure}[t]
	\begin{center}
		\includegraphics[width=0.5\columnwidth,keepaspectratio]%
		{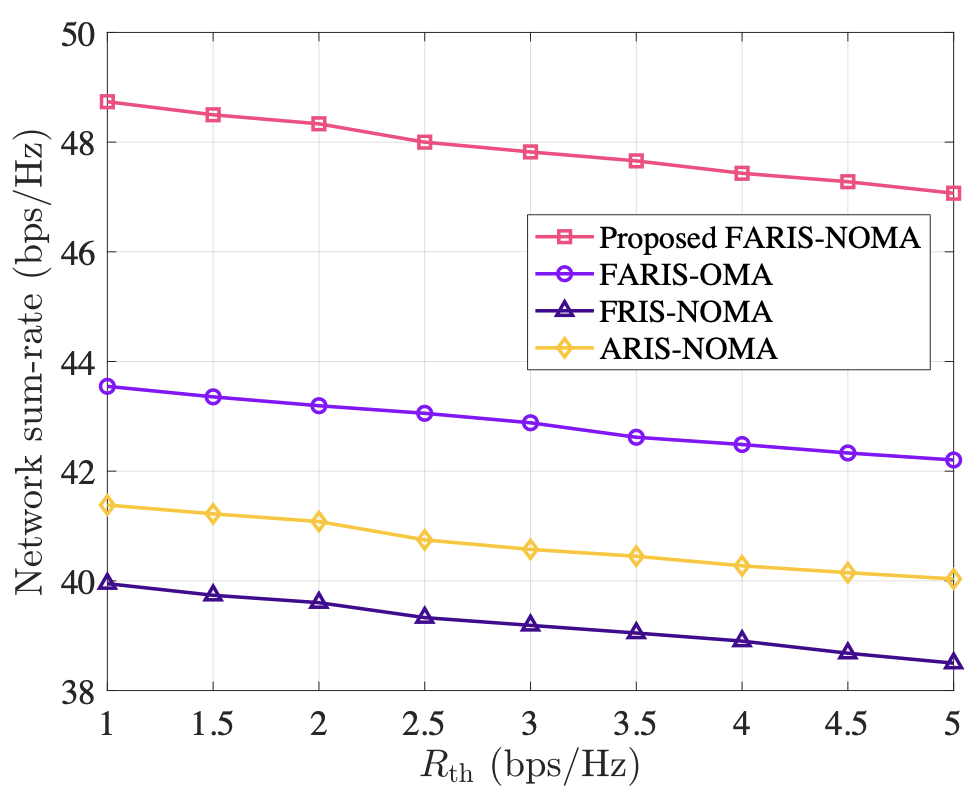}
\caption{Network sum-rate versus $R_{\mathrm{th}}$ for the proposed FARIS-NOMA and benchmarks.}
		\label{fig_rth}
	\end{center}
\end{figure}

Fig.~\ref{fig_rth} illustrates the network sum-rate versus $R_{\mathrm{th}}$. The sum-rate decreases for all strategies as $R_{\mathrm{th}}$ increases, since a more stringent QoS requirement restricts the feasible solution space in~\eqref{p0} and forces more resources to be allocated to satisfy the minimum rate constraints~\cite{RISEE}, thereby reducing the overall spectral efficiency. Nevertheless, the proposed FARIS-NOMA consistently achieves the highest performance; $11.7\sim22.4\%$ increase compared to baselines, as it can more effectively accommodate the stringent QoS requirements by efficiently allocating spatial and power resources.

\section{Conclusion}
In this paper, we developed a FARIS-aided framework for downlink NOMA that jointly leverages the spatial reconfigurability of fluid element selection and the amplification gain of active reflection. We formulated a network sum-rate optimization involving joint user clustering, power allocation, amplification design, discrete phase control, and element selection under QoS and power constraints. To address its intractability, we proposed a two-stage approach consisting of distance-based interleaved clustering and a per-cluster AO framework, where the clustering strategy efficiently induces channel disparity for effective SIC, and the AO design enables tractable optimization of the tightly coupled variables. The proposed design maintained algorithmic tractability while revealing a key insight: FARIS enhances NOMA performance by inducing favorable user disparity through fluid reconfiguration and compensating cascaded-path loss via active reflection. Simulation results demonstrated that the proposed FARIS-NOMA framework achieves near-optimal performance and consistently outperforms benchmarks, highlighting it as a practical solution for smart radio environments in 6G.

 \bibliographystyle{IEEEtran}
\bibliography{IEEEexample}

\end{document}